%% file: ms.tex
\documentclass[12pt,preprint]{aastex}

\newcommand{\degree}{$^{\circ}$}
\newcommand{\msol}{\hbox{$M_\odot$}}

\slugcomment{ApJS, in press}

\shorttitle{Point Sources GC Spitzer Survey}
\shortauthors{Ram\'{\i}rez et al.}

\begin{document}

\title{Point Sources from a {\it Spitzer} IRAC Survey of the Galactic Center}

\author{Solange V. Ram\'{\i}rez}
\affil{IPAC/Caltech, Pasadena, CA 91125}
\email{solange@ipac.caltech.edu}

\author{Richard G. Arendt}
\affil{CRESST/UMBC/GSFC, Code 665, Greenbelt, MD 20771}

\author{Kris Sellgren}
\affil{The Ohio State University, Columbus, OH 43210}

\author{Susan R. Stolovy}
\affil{{\it Spitzer} Science Center, Caltech, Pasadena, CA 91125}

\author{Angela Cotera}
\affil{SETI Institute, Mountain View, CA 94043}

\author{Howard A. Smith}
\affil{Harvard-Smithsonian Center for Astrophysics, Cambridge, MA 02138}

\author{Farhad Yusef-Zadeh}
\affil{Northwestern University, Evanston, IL 60208}

%%ABSTRACT
\begin{abstract}
We have obtained $Spitzer$/IRAC observations of the central
2.0\degree $\times$ 1.4\degree ($\sim$ 280 $\times$ 200 pc) 
of the Galaxy at 3.6\micron--8.0\micron.
A point source catalog of 1,065,565 objects is presented.
The catalog includes magnitudes for the point sources
at 3.6, 4.5, 5.8, and 8.0\micron , as well as $JHK_s$ 
photometry from 2MASS. 
The point source catalog is confusion limited with
average limits of 12.4, 12.1, 11.7, and 11.2 magnitudes
for [3.6], [4.5], [5.8], and [8.0], respectively.
We find that the confusion limits are spatially variable
because of stellar surface density, background surface brightness 
level, and extinction variations across the survey region.
The overall distribution of point source density with Galactic
latitude and longitude is essentially constant, but structure
does appear when sources of different magnitude ranges 
are selected.
Bright stars show a steep decreasing gradient with Galactic
latitude, and a slow decreasing gradient with Galactic
longitude, with a peak at the position of the Galactic center.
From IRAC color-magnitude and color-color diagrams,
we conclude that most of the point sources in our
catalog have IRAC magnitudes and colors characteristic
of red giant and AGB stars. 
\end{abstract}


%INTRODUCTION
\section{Introduction}

Our Galactic center (GC), at a distance of $\sim$ 8.0 kpc \citep{rei93}, 
is the closest galactic nucleus, observable at spatial 
resolutions unapproachable in other galaxies (1 pc$\approx$26\arcsec). 
The region has been intensely studied at wavelengths outside
the optical and UV regime, because it is unobservable with
optical telescopes due to obscuring dust in the Galactic plane.
The typical extinction toward the inner 200 pc 
is 25-30 visual magnitudes \citep{sch99,dut03}, and it is 
considerably higher towards molecular clouds located close to the GC.

The extent of the GC region is defined by a region of relatively 
high density molecular gas ($n_{H_2}$  $\sim$ $10^{4}$ cm$^{-3}$;
\citealt{bal87}), covering the inner 200 pc 
(170\arcmin $\times$ 40\arcmin, centered on the GC), 
called the Central Molecular Zone (CMZ). 
The CMZ produces 5\%-10\% of the Galaxy's infrared and Lyman 
continuum luminosity and contains 10\% of its molecular gas
\citep{bal87,mor96}.
The CMZ contains extremely dense giant molecular clouds 
\citep{mar04,oka05,bol06}, which are also very turbulent. 
Strong tidal shearing forces arise within the CMZ from a
gravitational potential well that increases as the
galactocentric radius decreases \citep{gus80},
culminating in the central black hole, Sgr A$^*$
\citep[e.g. ][]{sch03,ghe05}.

In the past, the study of the GC stellar population has been
concentrated primarily on the spatial regions surrounding 
three clusters of stars.
The Central Cluster contains the dense core of stars within
a few parsecs of the GC.  The cluster is composed of a mixture of
red supergiant and giant stars \citep[e.g. ][]{leb82,sel87} 
and young massive stars which exhibit
energetic winds as observed in their emission line spectra 
\citep[e.g. ][]{all90,kra91,lib95,blu95,tam96}. 
These bright, hot emission line stars trace an epoch of 
star formation that occurred about $10^7$ yr ago, while
the bright cool stars may be associated with either
the most recent epoch of star formation or older ones
\citep{hal92,kra95}. 
The separation of bright cool stars into
M supergiants (tracers of recent star formation) and 
less massive giants (tracers of older star formation)
has been used to study the star formation history
within the central cluster \citep{leb82,sel87,blu96,blu03}.
The Quintuplet and Arches Clusters are located at about
30 pc in projection from the GC.
Both clusters contain hundreds of massive O-B stars, 
and have ages of 2--4 Myr \citep{fig99}
These clusters are thought to be
the low-mass analog of the young ``super star clusters'' 
found in external galaxies \citep{all90,nag96,cot96,fig99}.

One stellar population that has been studied across a broader 
area ($\sim$200 pc) centered on the GC is the OH/IR stars
\citep[among others]{hab83,lin92,sjo98}.
OH/IR stars are oxygen rich Asymptotic Giant Branch (AGB) stars that 
are characterized by long period pulsations and high mass loss. 
Studies suggest that there are two distinct populations
of OH/IR stars observed towards the GC, which are
separated both spatially and kinematically \citep{lin92}.
The OH/IR stars that are more closely confined to the
Galactic plane and that have a net prograde rotational
velocity in the GC are also found to have higher OH maser expansion 
velocities than other OH/IR stars in the GC \citep{lin92}.
A higher expansion velocity requires
either that the star is more luminous than the average (thus a more massive
and younger star), or that it has a higher dust-to-gas ratio
(and thus a higher metallicity).

Four other infrared studies have surveyed areas within
200 pc including the CMZ. 
The 2MASS all sky survey \citep{skr06} and the DENIS survey
\citep{epc97} were limited by their wavelength range 
between 1.2 and 2.2\micron\, which was inadequate to characterize
the more obscured regions.
The {\it Midcourse Space Experiment} (MSX) 
observed between 6 and 25\micron\ and 
included the CMZ in its survey of the Galactic plane \citep{pri01}.
The angular resolution of MSX (18\arcsec\ at 8.3\micron), however,
was only sufficient to identify the brightest isolated individual stars.
Finally, portions of the CMZ were observed with ISOCAM as part 
of the ISOGAL survey, with $\sim$6\arcsec\ angular 
resolution at 7\micron\ and $\sim$13\arcsec\ angular
resolution at 15\micron\ \citep{omo03}.
The ISOGAL survey has been used to select young stellar object
(YSO) candidates in the GC \citep{fel02,sch06} within the 
restricted area coverage of the survey.

We have obtained $Spitzer$/IRAC observations of the central 
2.0\degree $\times$ 1.4\degree
($\sim$ 280 $\times$ 200 pc, including the CMZ) of the Galaxy at
3.6\micron--8\micron\ in Cycle 1 (GO 3677, PI: Stolovy).
These data represent the highest spatial resolution ($\sim$2\arcsec) 
and sensitivity uniform large-scale map made to date of the GC 
at mid-infrared wavelengths.
The IRAC data display complex filamentary structures in the interstellar
medium (S. Stolovy et al. 2007, in preparation) 
and allow us to detect optically obscured stellar sources.
In this paper, we present details on the data reduction and
point source extraction (Section \ref{sec_obs}). 
A catalog of the IRAC point sources band merged with 2MASS photometry
is presented in Section \ref{sec_cat}.
This catalog contains 1,065,565 point sources uniformly covering the CMZ.
The point source magnitude distributions are discussed in Section 
\ref{sec_num_den} and the point source distributions with Galactic
coordinates are examined in Section \ref{sec_gcoord}.
A discussion of the nature of the point sources in the catalog
is presented in Section \ref{sec_cmd_col}.
This is the first paper in an upcoming series on the $Spitzer$/IRAC 
observations of the Galactic center.

%OBSERVATIONS
\section{Observations and Data Processing \label{sec_obs}}

The InfraRed Array Camera \citep[IRAC,][]{faz04} on board the 
{\it Spitzer Space Telescope} \citep{wer04} was used to map the 
central regions of the Galaxy, with a spatial coverage of about 
2.0\degree \ in Galactic longitude by 1.4\degree \ in Galactic 
latitude. 
Details of the observations and data processing are given in 
S. Stolovy et al. (2007, in preparation) but we provide a brief summary here.

Each IRAC detector has a 5.2\arcmin $\times$ 5.2\arcmin \ field 
of view comprised of 256$\times$256 pixels and a mean pixel scale of 
1.22\arcsec\ per pixel. 
The four cameras have wavelengths of 3.6\micron, 4.5\micron, 
5.8\micron, and 8.0 \micron \  for Channels 1, 2, 3, and 4, 
respectively. 
Because all four cameras do not see the exact same region of sky
simultaneously and because of orientation constraints, a larger 
region was mapped to cover fully the desired central 
2.0\degree \ $\times$ 1.4\degree \  region. 
We used the shortest frame time (2 sec.) available for the full-array 
mode, corresponding to an on-source effective integration time of 1.2 
seconds per pixel. 
We took five dithered exposures (or frames) on the sky for each 
pointing, giving a total average 
on-sky integration time of 6 seconds. 
This dithering strategy allows us to correct for bad pixels, 
scattered light, and latent images and provided improved sampling 
of the point spread function. 
Additional processing as described in S. Stolovy et al. 
(2007, in preparation) was performed 
on the {\it Spitzer} Science Center (SSC) pipeline version S13.2 Basic 
Calibrated Data (BCD) products to
correct various artifacts (scattered light, latent images, column 
pulldown, and banding), producing much improved BCD frames and mosaics. 
One electronic artifact that was not corrected due to its non-linear nature
was the `bandwidth effect'.
This artifact causes extra `sources' to appear 4 
pixels away (and in some cases 8 pixels away) from very bright sources 
in Channels 3 and 4 along the readout direction (IRAC Data
Handbook, Version 3.0, Section 4.3.3). 
Thus, a few of these artifacts remain in our final mosaics.

Additional observations were taken in IRAC's sub-array mode for
areas in the survey that were affected by saturation.
These regions include the Central Cluster and the Quintuplet
cluster, plus 12 individual pointings.
In sub-array mode, a small section of the array is read out
(32 $\times$ 32 pixels = 40\arcsec\ $\times$ 40\arcsec),
and we used the shortest exposure time available of 0.02 sec.

\subsection{Mosaicking}

We used the SSC Mosaicking and Point source Extraction (MOPEX) package, 
version 030106 (available from the SSC web page 
\footnote{ssc.spitzer.caltech.edu/postbcd/download-mopex.html}) 
to create mosaics, extract point sources, and create source 
subtracted mosaics of the full-array data. 
MOPEX is composed of a series of PERL scripts. 
We used mosaic.pl to create mosaics, apex.pl to detect 
and measure fluxes of point sources, and apex\_qa.pl to create 
source subtracted images.
The script mosaic.pl performs interpolation and co-addition of FITS
images, with the additional functionality of detection of outliers.
The outliers are due to radiation hits, hot pixels, and bad pixels.
There are three outlier detection algorithms implemented within mosaic.pl.
The single frame outlier detection algorithm performs spatial filtering
within an individual BCD frame, flagging outliers above a user defined flux 
threshold, and below a user defined size. 
The multiframe outlier and the dual outlier detection algorithms determine
outlier pixels by stacking pixels taken in different exposures but at the same 
spatial location. 
The BCD frames are spatially matched by coincident point sources.
The multiframe outlier computes the statistics of the stacked pixels and 
finds the outliers above a user defined $\sigma$ threshold.
The dual outlier detection algorithm first detects all sources above a
user defined $\sigma$ threshold and then compares the number of 
detections for each spatial location to discriminate the outliers 
from the real sources.

The optimization of the parameters of the outlier detection algorithms
for our confusion-limited data was the most challenging part of the 
creation of the mosaics. 
The single frame outlier and the dual outlier algorithms flagged many 
real point sources as outliers, even using the most conservative set
of input parameters, mainly due to crowding of point sources in our images. 
Therefore, we adopted only the multiframe outlier detection algorithm 
with a conservative threshold of 50$\sigma$, 50$\sigma$, 60$\sigma$, and
25$\sigma$, for Channels 1, 2, 3, and 4 respectively, which gave satisfactory 
results in terms of the number of pixels flagged as outliers per BCD frame per 
unit of exposure time and the lack of real point sources flagged as outliers.
The expected number of radiation hits in one single IRAC frame is 
approximately 3 to 6 pixels per second in Channels 1 and 2, and 
approximately 4 to 8 pixels per second in Channels 3 and 4. 
Our observations comprise a total of 2,895 individual BCD frames per channel,
with an integration time of 1.2 s per frame. 
This predicts 
a total number of radiation hits of about 10,000-21,000 pixels 
in Channels 1 and 2, and about 14,000--28,000 pixels in Channels 3 and 4. 
The total numbers of flagged outliers determined by MOPEX using the
parameters described above were about 26,000, 26,000, 32,000, 
and 14,000 pixels in Channels 1, 2, 3, and 4, respectively, 
which are similar to expected values.

It is crucial to create a clean mosaic at each channel
before attempting to extract the point sources.
Our final mosaics represent a significant improvement over available 
SSC pipeline data products.
The top panels of Figures \ref{fig_ch1} and \ref{fig_ch4}
show the final full-array mosaics for Channels 1 and 4, covering a field of
view of 2.0\degree $\times$ 1.4\degree, centered on $l$=0,0 and $b$=0.0
(see S. Stolovy et al. 2007, in preparation, for final mosaics including 
the sub-array observations).

\subsection{Source extraction from full-array data \label{full_extr}}

The source extraction was performed using the MOPEX script apex.pl, set up
so it utilizes data products created with the MOPEX script mosaic.pl, in 
particular those concerning outlier detection. 
The script apex.pl performs the source detection in a background subtracted
mosaic.
It provides two measurements of the flux: one comes from a point response
function (PRF) fitting and the other comes from an aperture measurement.

The PRF is the telescope point spread function convolved with the instrument
response function. 
The algorithm that fits the PRF to the BCD sources allows a determination of
the local background, which is advisable to use in crowded fields
with variable background level such as those observed in the GC.
Note that the flux uncertainty and the signal to noise ratio (SNR) 
as provided by apex.pl are the uncertainty and SNR
of the PRF fitting algorithm.
The PRFs used in the flux measurement were provided by the SSC.

The aperture flux measurement is performed on the mosaic image.
The usual background estimates include performing the aperture 
measurement on a median filtered mosaic or using an annulus around
the detected sources. 
Neither of these two methods for estimating the background are 
appropriate for our survey due to point source crowding and 
variable and high background levels. 
Instead, we used the local background determined by the PRF fitting to
subtract the background contribution to the corresponding aperture flux.
We measured the flux within a small aperture of 2 pixels radius 
(about 2.44\arcsec) to avoid confusion, then subtracted the background
contribution, and finally applied the corresponding aperture correction.
The value of the aperture corrections are
1.213, 1.234, 1.379, and 1.584 for Channels 1, 2, 3, and 4, respectively, 
as provided in the IRAC Data Handbook, Version 3.0. 
The script apex.pl does not provide a measure of the uncertainty of 
the aperture flux. 
We estimate the aperture flux uncertainty by performing an aperture 
measurement of the same size as the photometric aperture directly in the 
mosaic of the uncertainty images (data product of mosaic.pl). 

The source extraction was performed in each of the 12 AORs separately
because of the lack of enough computing memory to process the entire
data set simultaneously. 
The resulting source lists for each AOR were combined to obtain a 
total source list for the whole survey.
During this process we also rejected sources that were within a certain 
radius from a bright source ($\sigma$ threshold $>$ 30) and sources that 
were detected on top of extended emission in Channels 3 and 4. 
The radius of avoidance was 4.5\arcsec, 5.8\arcsec, 6.4\arcsec, and 
8.3\arcsec \ for Channels 1, 2, 3, and 4, respectively, which 
corresponds roughly to the radius of the second minimum of the diffraction 
pattern of the PRF.
Note that rejecting faint sources within the radius of avoidance 
will also discard possible artifacts due to the 4-pixel bandwidth
effect, present in Channels 3 and 4.
The `bandwidth effect' causes extra `sources' to appear 4 pixels away 
(and in some cases 8 pixels away) from very bright sources
in Channels 3 and 4 along the readout direction (IRAC Data
Handbook, Version 3.0, Section 4.3.3).
It is possible that some saturated sources were misidentified as many 
individual sources, each of them too faint to trigger the avoidance radius.
To discriminate between a point source and an extended source, we use the fact 
that a point source should have the same aperture corrected flux independent 
of the size of the aperture used.
We flagged a source as extended if the aperture corrected flux from a 3 pixel 
radius aperture differs by more than 15\% from a 2 pixel radius aperture 
corrected flux. 
Sources flagged as extended are not included in the catalog.

We found that the PRF fluxes and aperture fluxes
agreed to within 12\% overall.  We did, however, find that
the PRF fluxes were systematically lower than the aperture 
fluxes by 13-12\% for Channels 1 and 2, and higher by
7\% for Channel 4.  No significant difference was found
for Channel 3. This difference is likely to arise from errors 
in PRF normalization. We measured the difference between the
aperture fluxes and the PRF fluxes by first determining
the IRAC colors of foreground sources with low amounts
of reddening.
Most of the foreground stars are expected to be red giant stars,
whose IRAC colors should be near zero (M. Cohen, private communication;
IRAC Handbook).
\citet{sch99} and \citet{dut03} have determined extinction maps at
the Galactic center distance.
The minimum extinction observed for a source located at the distance of 
the GC is $A_K$ = 1 magnitudes \citep{sch99,dut03}.
We therefore selected as foreground stars those sources with 
$A_K <$ 1 or $(J-K_s) <$ 1.5.
There are 6816 sources in our source list that have $(J-K) < $1.5,
and their mean IRAC colors are listed in Table \ref{tab1}, for
both PRF and aperture magnitudes.
Aperture IRAC colors are closer to zero than PRF IRAC colors.
The PRF method, however, is generally superior than aperture photometry
in crowded fields, and also shows less scatter at fainter magnitudes.
We adjusted the PRF fluxes for Channels 1, 2, and 4 such that the 
median ratio of the two extraction methods was 1. 
The multiplicative factors applied to the PRF fluxes were 1.13, 
1.12, 1.00, and 0.93 for Channels 1, 2, 3, and 4 respectively.
The IRAC colors of the final photometry are also listed in 
Table \ref{tab1}.

We produced point source-subtracted images, using the MOPEX script apex\_qa.pl, 
to assess the effectiveness of the extraction and to compare the flux results 
obtained with the PRF fitting and aperture photometry. 
We found that the PRF fitting occasionally failed, producing a flux that 
was much too high. 
For the cases where the adjusted-PRF/aperture flux ratio exceeded 1.5, 
we adopted the aperture value of the flux and its corresponding 
uncertainty was derived as described above.
The source-subtracted residual images show even fainter sources but 
we did not attempt to extract them. 
Additionally, the brighter sources close to saturation are in the 
nonlinear regime and therefore do not match the shape of the PRF.
We did not attempt to subtract highly saturated sources. 

\begin{figure}
\epsscale{.75}
\plotone{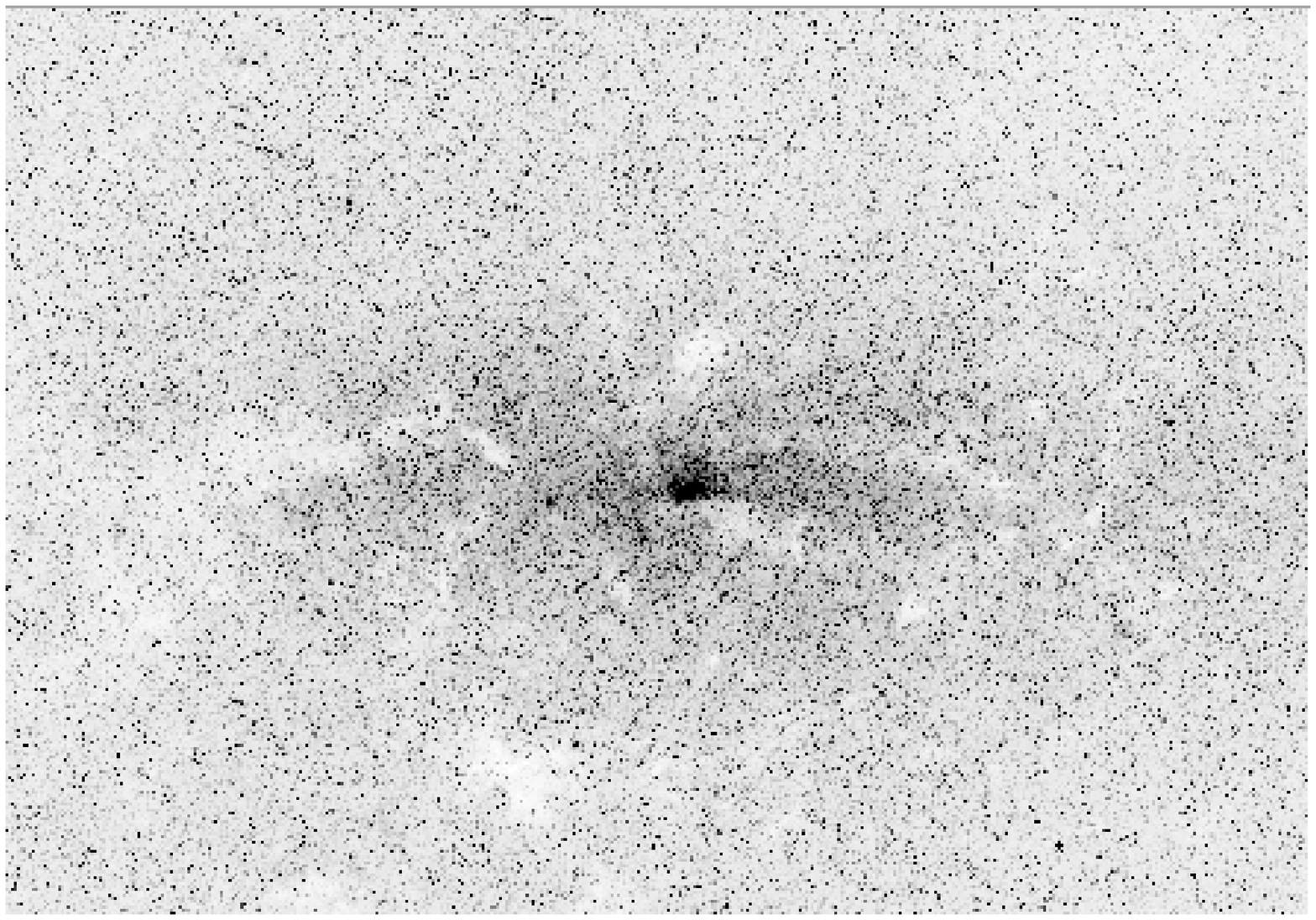}\\
~\\
~\plotone{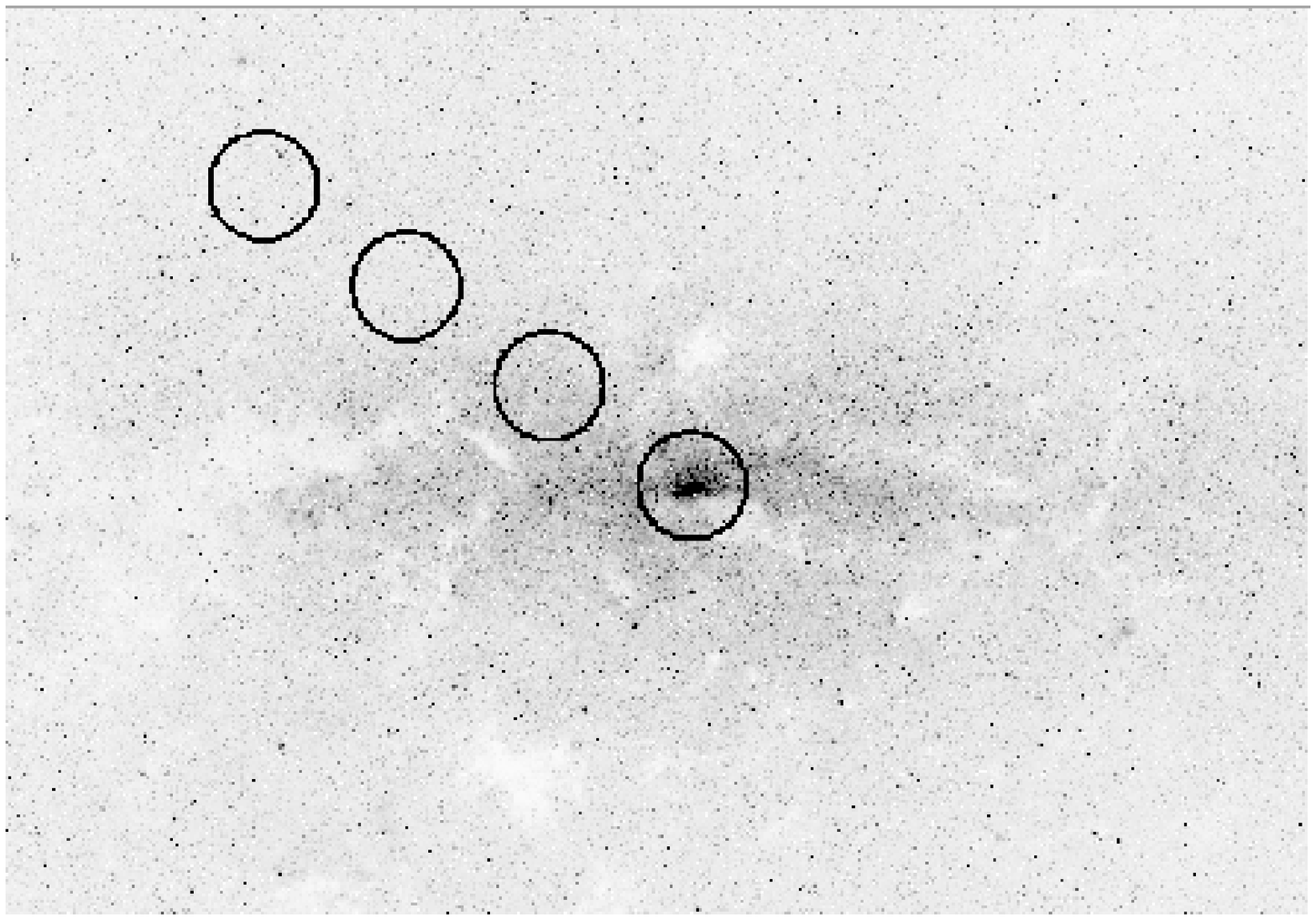}
\epsscale{1.0}
\figcaption{
IRAC Channel 1 (3.6 \micron) mosaic ({\it top panel}) and source subtracted 
mosaic ({\it bottom panel}) of the Galactic center, covering an area of 
2.0\degree\ $\times$ 1.4\degree, centered on $l$=0.0, $b$=0.0
(Galactic north is up, Galactic east is to the left).
The mosaics are shown in reverse grayscale with the same scale.
The circular areas shown in the {\it bottom panel} are centered on 
$l$=359.946, $b=-0.0378$; $l$=0.166, $b$=0.1162; 
$l$=0.386, $b$=0.2702; and $l$=0.606, $b$=0.4242.
These circular areas are used in this paper to study the distribution of point
sources in locations with different source densities.
\label{fig_ch1}}
\end{figure}

\begin{figure}
\epsscale{.75}
\plotone{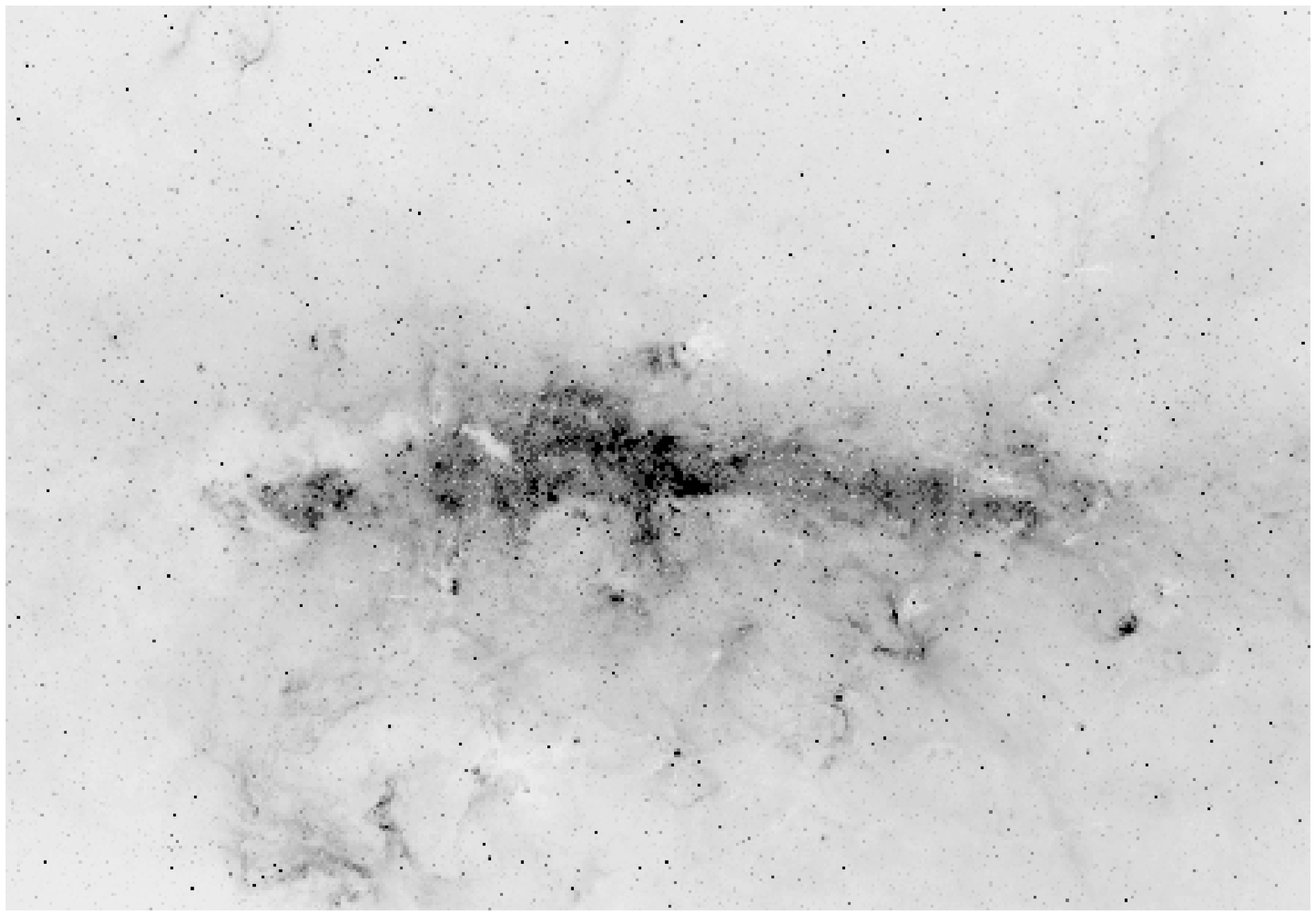}\\
~\\
~\plotone{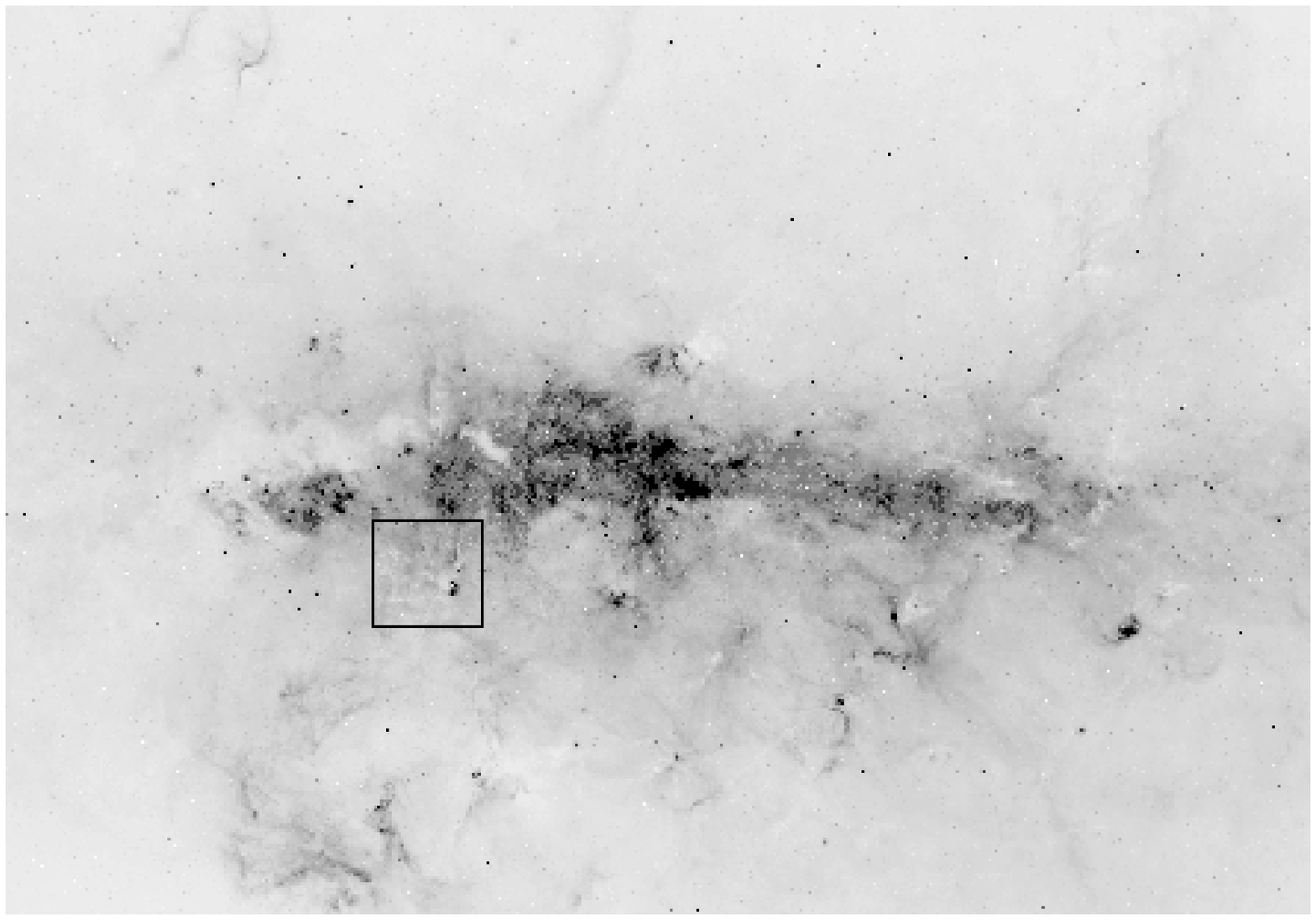}
\epsscale{1.0}
\figcaption{
IRAC Channel 4 (8 \micron) mosaic ({\it top panel}) and source subtracted 
mosaic ({\it bottom panel}) of the Galactic center, covering an area of 
2.0\degree$\times$1.4\degree, centered on $l$=0.0, $b$=0.0
(Galactic north is up, Galactic east is to the left).
The mosaics are shown in reverse grayscale with the same scale.
The box plotted in the {\it bottom panel} shows the location of the blown up
section detailed on Figure \ref{fig_det}.
\label{fig_ch4}}
\end{figure}

The source subtracted mosaics for Channels 1 and 4 are shown in 
the bottom panels of Figures \ref{fig_ch1} and \ref{fig_ch4}.
The four circular areas shown in the bottom panel of Figure \ref{fig_ch1}
will be used in Section \ref{sec_num_den} to study the distribution of
point sources in different locations within our field of view.
The circular areas have a radius of 5\arcmin\ and they are centered on 
$l$=359.946, $b=-0.0378$; $l$=0.166, $b$=0.1162; 
$l$=0.386, $b$=0.2702; and $l$=0.606, $b$=0.4242.
 
\begin{figure}
\epsscale{.50}
\plotone{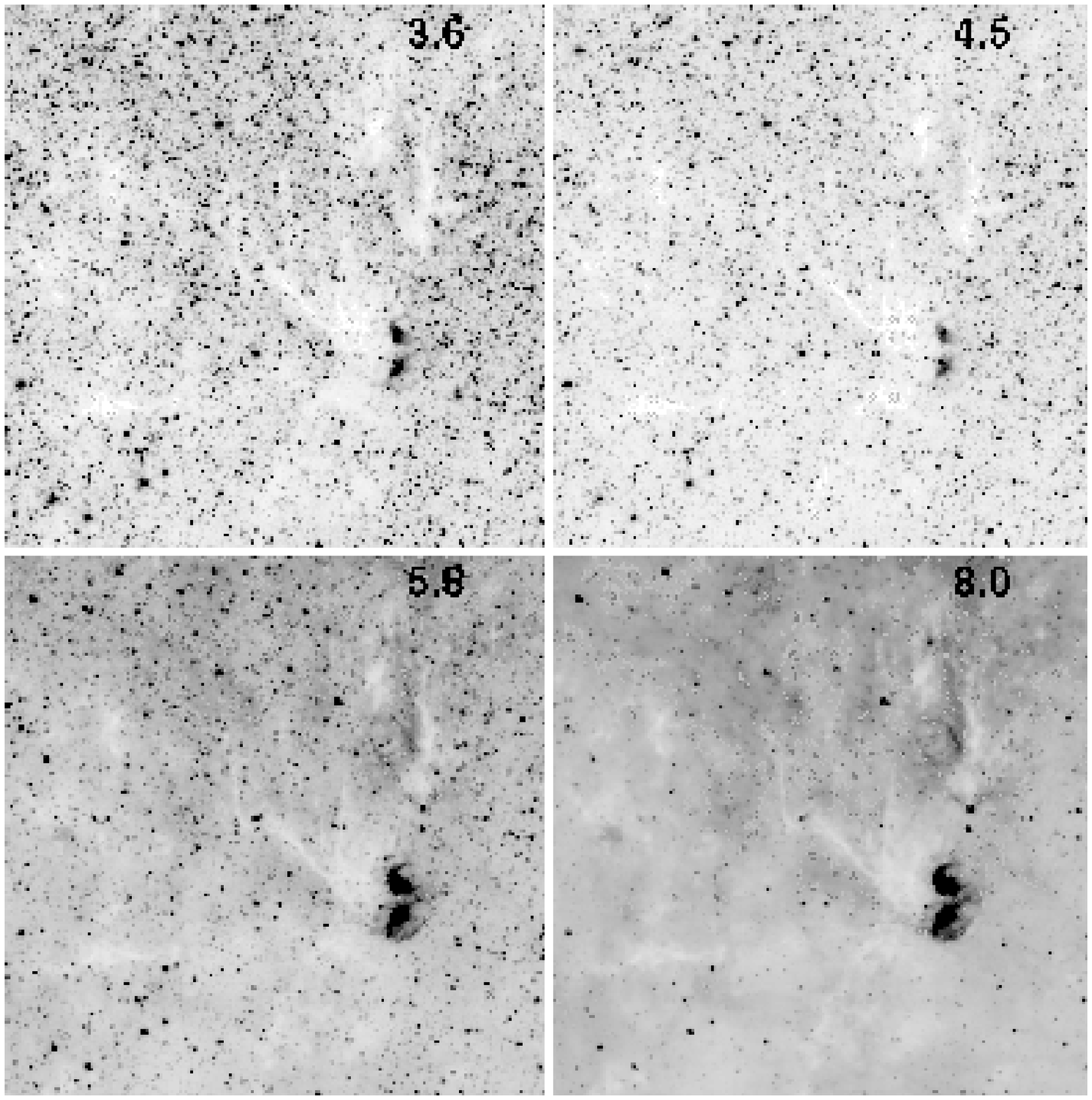}\\
~\\
~\plotone{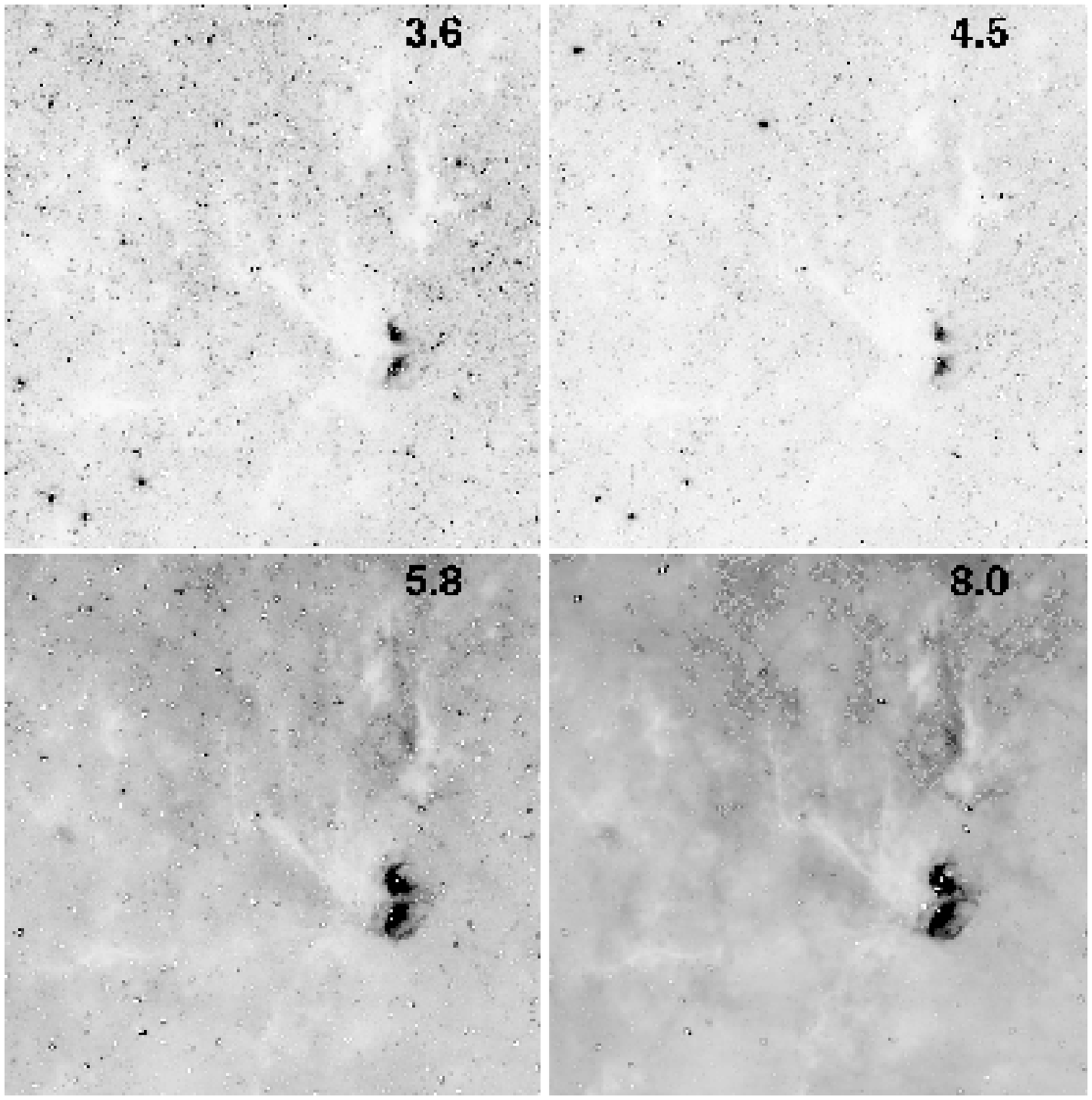}
\epsscale{1.0}
\figcaption{Detail of the IRAC mosaics and source subtracted mosaics
showing 10\arcmin\ $\times$ 10\arcmin\ field of view,
centered on ($l$=0.3523, $b$=$-0.17427$).
The original mosaics are shown in the top set of four panels and the
corresponding source subtracted images are shown in the bottom set. 
Each IRAC channel is labeled in the top right corner of the 
individual images.
Note the differences in source densities and extended emission among the 
different IRAC channels.
The residuals from the point sources are larger in Channel 1, and smaller in
Channel 4, because the PRF is better sampled at longer wavelengths.
\label{fig_det}}
\end{figure}

Figure \ref{fig_det} shows a 10\arcmin $\times$ 10\arcmin\ field of view
centered on ($l$=0.3523, $b$=$-0.17427$), marked as a box
in the bottom panel of Figure \ref{fig_ch4}.
Figure \ref{fig_det} shows the differences in source densities and 
extended emission among the different IRAC channels.
Residuals in Channel 4 are the smallest because the PRF is better sampled than 
in the other channels.

The total number of sources detected at a level of 3$\sigma$ or above
in each channel was: 735,020, 700,923, 493,207, and 323,512 for Channels
1, 2, 3, and 4, respectively.
All the sources detected and measured by MOPEX at the 3$\sigma$ level 
are listed in Table \ref{tab2}, \ref{tab3}, \ref{tab4}, and \ref{tab5} 
(shown partially, available entirely in the electronic version).
The columns of Table \ref{tab2}--\ref{tab5} are as follows: 
Source Identification, IRAC Channel, Position (Equatorial and Galactic), 
Flux in mJy, Flux uncertainty in mJy, Number of Observations (BCD frames 
used in the Flux measurement), Signal-to-noise ratio, and Flux Method 
as explained in the previous section.

\begin{figure}
\plotone{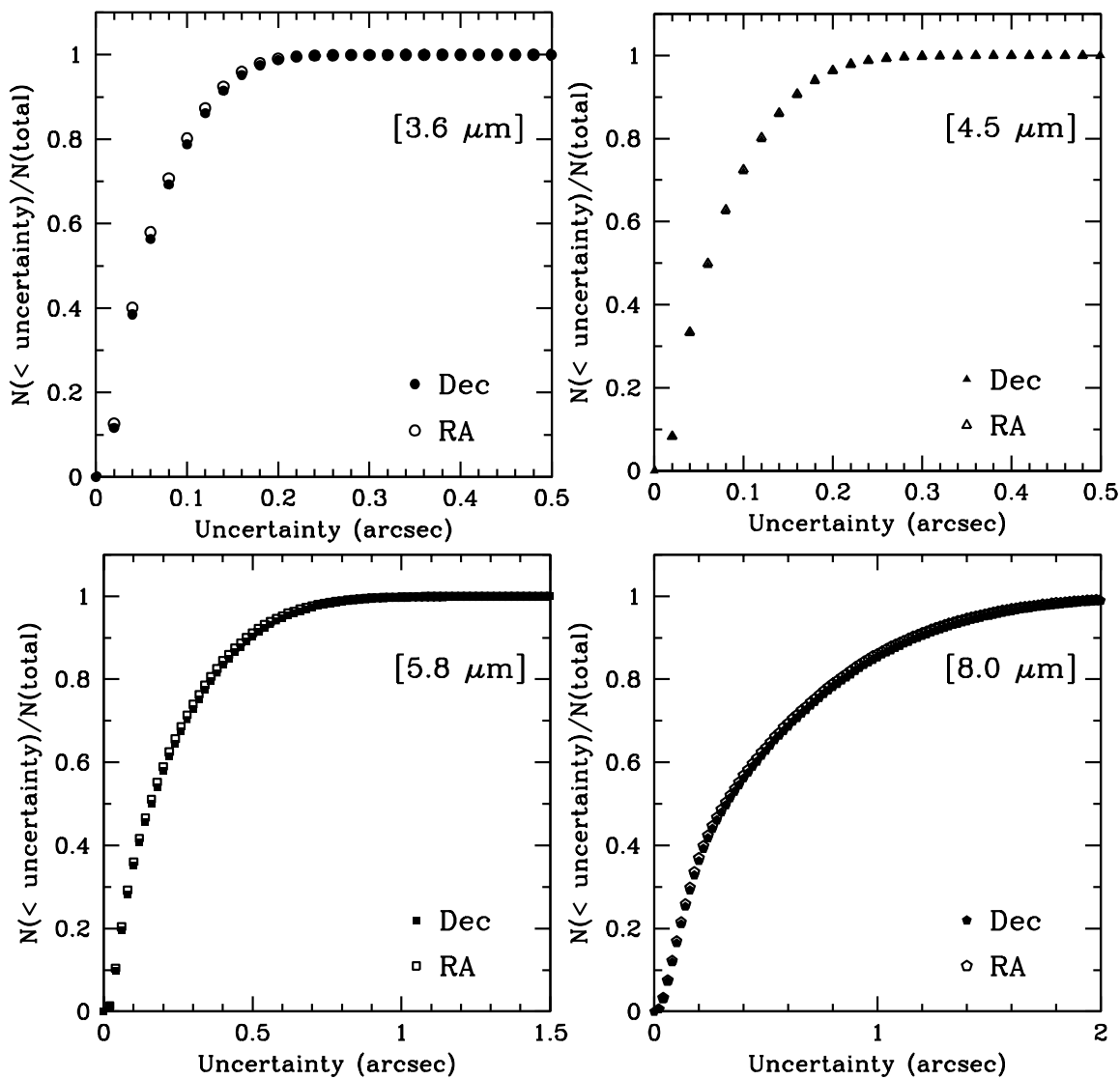}
\figcaption{Cumulative distribution of positional uncertainties.
Open symbols denote the cumulative distribution of positional 
uncertainties in right ascencsion and the filled symbols denote the cumulative 
distribution of positional uncertainties in 
declination, both in units of arcsec.
All sources observed in each IRAC channel are included in these
cumulative distributions.
\label{fig_pos_unc}}
\end{figure}

The cumulative distribution of positional uncertainties is shown 
in Figure \ref{fig_pos_unc}, 
for both right ascension (open symbols)
and declination (filled symbols).
The overall distributions of positional uncertainties are
similar between right ascension and declination,
for all four IRAC channels.
We found that 90\% of the sources in our survey have positional 
uncertainties less than 0.13\arcsec, 0.16\arcsec, 0.48\arcsec, 
and 1.18\arcsec \ for Channels 1, 2, 3, and 4, respectively,
corresponding to the typical positional uncertainties in our survey.
Also, 99\%, 96\%, 59\%, and 36\% of the Channel 1, 2, 3, and 4
sources, respectively, have a positional uncertainty less than 
0.2\arcsec.
In the process of merging the IRAC point source lists with 2MASS
(see details below), we found systematic offsets between IRAC positions
in different channels and between the IRAC and 2MASS positions. 
These offsets are $\sim -$0.25\arcsec \ in 
right ascension, and $\sim$0.15\arcsec \ in declination,
and have been applied to Ch. 1-4 such that the IRAC astrometry 
in all channels should 
now match the 2MASS astrometry in our final catalog.

\begin{figure}
\plotone{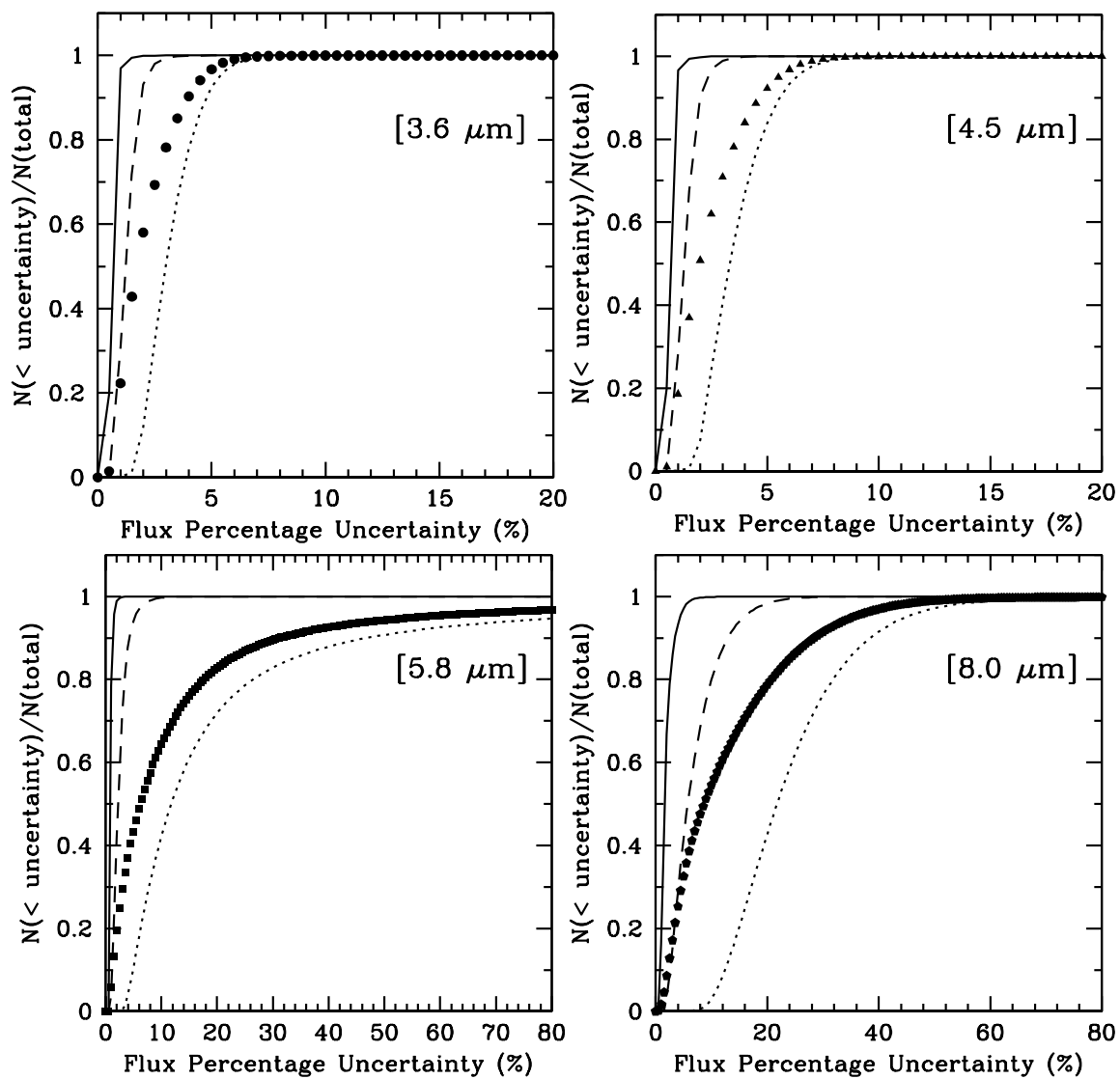}
\figcaption{Cumulative distribution of flux uncertainties.
All sources observed in each IRAC channel are included in the
cumulative distributions, plotted with filled symbols. 
The $solid$, $dashed$, and $dotted$ $lines$ show the
cumulative distribution of flux uncertainties for sources of
three different source brightness ranges, respectively bright, medium, and
faint.
\label{fig_flux_unc}}
\end{figure}

Figure \ref{fig_flux_unc} shows the cumulative distribution of percentage 
flux uncertainty, derived as described above.
The cumulative distribution of the percentage flux uncertainty for the
four IRAC channels is shown in separate panels.
The open symbols denote the cumulative distribution of percentage flux
uncertainties for all the sources in each of the IRAC channels.
We found that 90\% of the sources in our survey have a percentage
flux uncertainty less than 4.0\%, 5.0\%, 31\% and 28\%  for
Channels 1, 2, 3, and 4, respectively, corresponding to the typical 
percentage flux uncertainties in our survey.
Also, 99\%, 99\%, 64\%, 56\% of the Channel 1, 2, 3, and 4
sources, respectively, have a percentage flux uncertainty less than 10\%.
The solid, dashed, and dotted lines in Figure \ref{fig_flux_unc} 
show the cumulative distribution of percentage flux uncertainties for 
sources of 
three different source brightness ranges, {\it bright}, {\it medium}, and 
{\it faint}, respectively, as defined in Section \ref{sec_num_den}. 
The distribution of percentage flux uncertainties for Channels 3 and 4 is
dominated by the distribution of percentage flux uncertainties for faint
sources. 
Channels 3 and 4 mosaics show a wide range in background levels
on top of which faint sources are measured. 
Variations in the local background may be the cause of the larger
uncertainties in the flux measured in those channels.

\subsection{Source extraction from sub-array data}

In order to recover useful photometry from the small saturated
regions in the full-array observations, we performed photometry
on sub-array data, which consists of mosaics of the Central
Cluster (SgrA) and the Quintuplet Cluster, plus 12 individual 
pointings.
We used the interactive IDL program xstarfinder \citep{dio00},
because the parameters used for the full-array data using MOPEX were
not appropriate for the small sub-array observations.
We also tested photometry using the IRAF source extraction program
``daophot'' but found that daophot gave less reliable results than 
xstarfinder.
 
The PRF was constructed from a composite of sub-array observations of well
exposed, isolated single sources.
These sources were chosen from the twelve individual pointing observations, 
excluding observations with higher than typical noise or
with other stars within a $\sim$10\arcsec\ radius of the main source.
PRF's were made using both xstarfinder and daophot and it was determined
that the point source subtracted residual images were superior for the
daophot PRFs; thus, the PRFs from daophot were used in the source extraction.

For each input mosaic, the surface brightness error per pixel was
computed from each input mosaic directly.
This error is photon-noise-dominated and well fit with a gaussian distribution.
The flux errors for the extracted point sources are statistical only and
do not reflect differences in the flux estimate that may arise from
methodology, e.g., using a different set of extraction parameters such
as the background smoothing box.
We expect that the systematic errors may
exceed the quoted random errors.
The source extraction computes a correlation factor, which is a measure
of the goodness of fit to the PRF, with 1.0 being a perfect fit.
The correlation factor for all extracted sources was 0.75 at minimum,
but most sources had a factor exceeding 0.9.
A median smoothing size of 7 times the FWHM was used for the background
determination.

Table \ref{tab6} lists the sub-array photometric results including 13
sources in the dozen individual pointings or ``sat''  fields, 104 sources
located in the Central Cluster or ``sgra'' field, and 90 sources 
located in the Quintuplet Cluster or ``quint'' field.
The sub-array source table lists the brightness and its uncertainty
in magnitudes for each IRAC Channel.

%RESULTS

\section{Catalog of Point Sources in the Galactic Center \label{sec_cat}}

We bandmerged our IRAC full-array source list for 
each channel with the sources 
in the 2MASS catalog \citep{skr06} located in the same field of view 
as our IRAC observations. 
The merging procedure was done as follows: We first matched and merged,
via a positional association, Channel 1 and 2, then Channel 3, then 
Channel 4, and finally 2MASS. 
A match was defined as the closest counterpart within a 1\arcsec\ radius.
The position used for the final merged list was always that corresponding 
to the shortest IRAC wavelength at which a source is seen.
The sub-array photometry was incorporated into the band merged list.
The full-array photometry of each of the 13 sources in the ``sat'' 
fields was individually replaced by the sub-array photometry.
All the sources with full-array photometry located within 
40\arcsec\ of the Central Cluster 
(RA=17 45 40.0, DEC=$-29$ 00 28) and within 
42\arcsec\ of the Quintuplet Cluster 
(RA=17 46 16.1, DEC=$-28$ 53 43) were discarded. 
The sub-array photometry of the ``sgra'' and ``quint'' fields were added
to the band merged list.

The merged source list was further studied and additional flags 
regarding the detection reliability, position comments, and
photometric quality were added.
Magnitudes for each of the IRAC channels were computed using 
zero point fluxes of 280.9 Jy, 179.7 Jy, 115.0 Jy, and 64.13 Jy,
for IRAC Channels 1, 2, 3, and 4 respectively, as provided by
\citet{rea05}.

\begin{figure}
\plotone{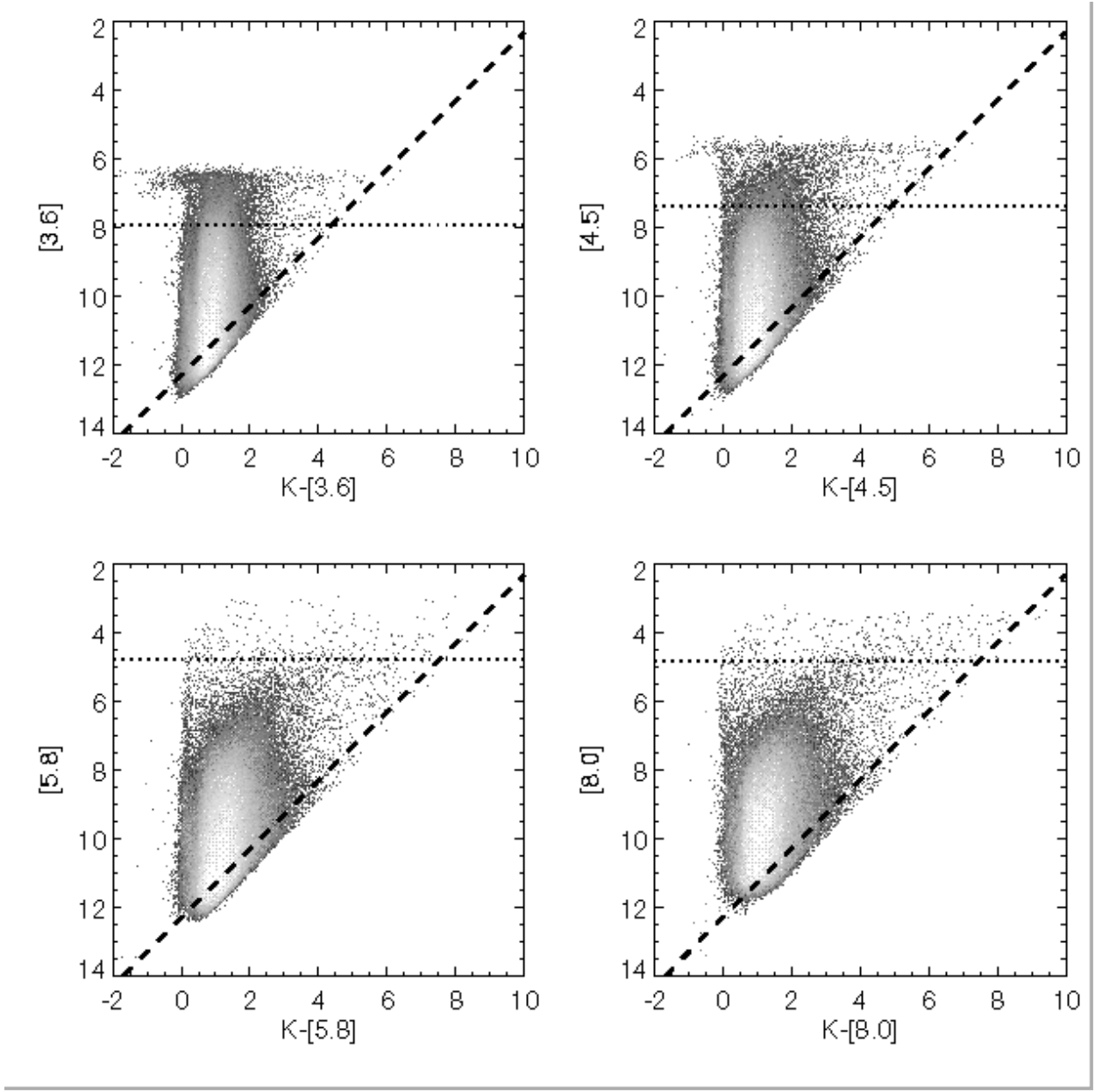}
\figcaption{Color-magnitude diagrams (CMDs), 
or plots of the difference between $K_s$ and IRAC magnitude vs. 
IRAC magnitude.
The CMD for the four IRAC magnitudes are shown in the corresponding
panels.
Only non-saturated high quality IRAC and $K_s$ magnitudes (IRAC magnitudes
with SNR$>$10 and 2MASS photometric quality flag equal to `A' (SNR$>$10)) 
are included in this figure.
The gray scale shows the number density distribution of
sources, with white being the highest density.
The {\it dotted line} shows the magnitude corresponding to the saturation
fluxes of 190 mJy, 200 mJy, 1400 mJy, and 740 mJy 
(7.92, 7.38, 4.79, and 4.84 magnitudes) for IRAC Channels
1, 2, 3, and 4 respectively, as provided by the $Spitzer$
Observer's Manual (SOM).
The {\it dashed lines} corresponds to the completeness limit of $K_s$=12.3
magnitudes of the 2MASS point source catalog within a 6\degree \ radius
of the Galactic center.
\label{fig_sat}}
\end{figure}

One of the qualities studied in the band merged source list was
flux saturation for the full-array photometry. 
We needed to explore whether the saturation values for point sources 
provided by the IRAC documentation were appropriate for our survey.
Figure \ref{fig_sat} shows color magnitude diagrams (CMDs) using each 
IRAC magnitude and the $K_s$ magnitude from 2MASS
\citep{skr06}. 
Only non-saturated $K_s$ magnitudes with 2MASS photometric quality 
flags (ph\_qual) equal to `A' (SNR$>$10) are included in Figure \ref{fig_sat}.
The gray scale shows the number density distribution of
sources, with white being the highest density.
Any anomalies in the bright regime of the CMDs
are due to non-linear and saturation effects in the IRAC magnitudes.
The horizontal dotted lines show the magnitudes 
corresponding to the saturation 
fluxes of 190 mJy, 200 mJy, 1400 mJy, and 740 mJy 
(7.92, 7.38, 4.79, and 4.84 magnitudes) for IRAC Channels 1, 2, 3, 
and 4 respectively, as provided by the $Spitzer$ Observer's Manual, 
Version 7.1.
The slanted dashed lines correspond to the completeness limit of $K_s$=12.3
for the 2MASS point source catalog within a 6\degree \ radius
of the Galactic center \citep[see Sec IV.7, ][]{cut03}. 
Figure \ref{fig_sat} also demonstrates that the saturation fluxes 
provided by the $Spitzer$ Observer's Manual (Version 7.1) are indeed
appropriately applied to our survey, as can be seen from the 
anvil-shaped tops of the CMDs for Channels 1 and 2.
Note that the point source fluxes are superimposed on a high
background, and the combination of both is likely to explain 
the saturation level appearing to be conservative when it really is 
appropriate.
Saturated sources are retained in our point source catalog, as long as
they are recognized as a point source by the apex.pl script.
If the flux of a source is greater than the saturation flux
provided by the $Spitzer$ Observer's Manual, 
then all its measured quantities are kept in the catalog, but
the flux flag
of that source in that channel is set to `3' in our final catalog.

The coverage of our survey has some incompleteness due to the fact
that the four IRAC cameras do not see exactly the same region of the sky. 
The coverage for each source at each channel was determined by
measuring the value of the pipeline coverage map at the position
of each source. 
The coverage value is the same as the number of available BCD frames at 
the position of each source, and it is also listed in our
final catalog for each channel.
The location of each of the sources is flagged in our catalog
by the Position flag. 
If a source is located in the area of incomplete coverage
(coverage value is equal to zero in at least one channel),
then the Position flag is set to `0'.
Sub-array photometry has been incorporated into the catalog,
in particular in the location near the
Central Cluster and near the Quintuplet Cluster.
If a source is located within 40\arcsec \ of the Central Cluster
(RA=17 45 40.0, DEC=$-29$ 00 28), the Position flag is set to `2'.
If a source is located within 42\arcsec \ of the Quintuplet Cluster
(RA=17 46 16.1, DEC=$-28$ 53 43), the Position flag is set to `3'.
The sources with Position flag set to `2' and `3' have photometry 
from the sub-array observations.

The reliability of the sources in our band merged list can
be estimated by applying the ``2+1" criterion as defined
by the Galactic Legacy Infrared Mid-Plane Survey Extraordinaire
(GLIMPSE) v1.5 Data Products Description (available from 
the GLIMPSE Documents web page 
\footnote{http://www.astro.wisc.edu/sirtf/docs.html}).
The $M/N$ ratio is defined as the ratio between $M$ number of
detections over $N$ number of possible observations
(coverage value).
The ``2+1" criterion requires $M/N\geq0.6$ in one IRAC
band, $M/N\geq0.4$ in an adjacent band.
Sources that satisfy the GLIMPSE ``2+1" criterion have the
``2+1" flag set to `1' in our final catalog.

Our final catalog of point sources in the Galactic center is
listed in Table \ref{tab7}.
The columns of our point source catalog are explained as follows:

\begin{description}

\item[Column 1, Source ID:] Designation of the detected source.

\item[Column 2, R.A.:] Right Ascension in J2000.

\item[Column 3, Dec.:] Declination in J2000.

\item[Column 4, $l$:] Galactic Longitude.

\item[Column 5, $b$:] Galactic Latitude.

\item[Column 6, ``2+1" Flag:] Set to `1' when the source 
satisfies the GLIMPSE ``2+1" criterion ($M/N\geq0.6$ in one IRAC
band, $M/N\geq0.4$ in an adjacent band), set to `0' otherwise.

\item[Column 7, Pos. Flag.:] Set to `0' when the source
is located in areas of incomplete coverage, set to `2'
when the source is within 40\arcsec \ of the Central Cluster, 
set to `3' when the source is within 42\arcsec \ of the Quintuplet 
Cluster, otherwise is set to `1'.

\item[Column 8, 2MASS ID:] 2MASS identification number from the 2MASS Catalog,
set to `none' when there is no 2MASS counterpart.

\item[Column 9, $J$:] $J$ magnitude from the 2MASS Catalog, set to `$-9.999$'
when not detected.

\item[Column 10, $J$ unc.:] Uncertainty of $J$ magnitude from the 2MASS 
Catalog, set to `$-9.999$' when not measured.

\item[Column 11, $H$:] $H$ magnitude from the 2MASS Catalog, set to `$-9.999$'
when not detected.

\item[Column 12, $H$ unc.:] Uncertainty of $H$ magnitude from the 2MASS 
Catalog, set to `$-9.999$' when not measured.

\item[Column 13, $K_s$:] $K_s$ magnitude from the 2MASS Catalog, set to 
`$-9.999$' when not detected.

\item[Column 14, $K_s$ unc.:] Uncertainty of $K_s$ magnitude from the 
2MASS Catalog, set to `$-9.999$' when not measured.

\item[Column 15, Qual. Flag:] Photometric quality flag (ph\_qual) 
from the 2MASS Catalog. 
It is composed of three letters, one for each 2MASS filter ($JHK_s$). 
The letters can be: 
A (SNR $>$ 10), B (SNR $>$ 7), C (SNR $>$ 5), D (no SNR requirement),
E (poor profile-fit photometry), 
F (detection without photometric uncertainty), 
U (detection with upper limit on magnitude), X (detection without 
brightness estimate). Set to `ZZZ' when there is no 2MASS counterpart.

\item[Column 16, ch1 ID:] Channel 1 identification number 
from point source list, also listed in Column 1 of Table \ref{tab2},
set to `none' when there is no Channel 1 counterpart.

\item[{Column 17, [3.6]:}] Channel 1 magnitude, computed using the Flux from 
Column 7 of Table \ref{tab2}, and the corresponding zero point flux, 
set to `$-9.999$' when not detected in this IRAC channel.

\item[{Column 18, [3.6] unc.:}] Uncertainty of Channel 1 magnitude, 
computed using the Flux and its uncertainty from Columns 7 and 8 of 
Table \ref{tab2}, and the corresponding zero point flux, 
set to `$-9.999$' when not detected in this IRAC channel.
 
\item[{Column 19, [3.6] SNR:}] Signal to Noise ratio of Channel 1 magnitude.
set to `$-9.9$' when not detected in this IRAC channel.

\item[Column 20, ch1 Flag:] Channel 1 magnitude flag. 
Set to `1' when flux and hence magnitude comes from PRF fitting algorithm, 
set to `2' when flux and hence magnitude comes from aperture corrected 
measurement, 
set to `3' when the full-array flux is greater than corresponding saturation 
limit,
set to `4' when the photometry comes from the sub-array observations,
set to `0' when there is no detection in this IRAC channel.

\item[Column 21, ch1 Cov.:] Number of available BCD frames at the 
position of the source for the corresponding IRAC channel ($N$).

\item[Column 22, ch1 $M/N$:] Ratio between $M$ number of detections 
over $N$ number of possible observations (ch1 Cov.).

\item[Column 23, ch2 ID:] Same as Column 16, but for Channel 2.

\item[{Column 24, [4.5]:}] Same as Column 17, but for Channel 2.

\item[{Column 25, [4.5] unc.:}] Same as Column 18, but for Channel 2.

\item[{Column 26, [4.5] SNR:}] Same as Column 19, but for Channel 2.

\item[Column 27, ch2 Flag:] Same as Column 20, but for Channel 2.

\item[Column 28, ch2 Cov.:] Same as Column 21, but for Channel 2.

\item[Column 29, ch2 $M/N$:] Same as Column 22, but for Channel 2.

\item[Column 30, ch3 ID:] Same as Column 16, but for Channel 3.

\item[{Column 31, [5.8]:}] Same as Column 17, but for Channel 3.

\item[{Column 32, [5.8] unc.:}] Same as Column 18, but for channel 3.

\item[{Column 33, [5.8] SNR:}] Same as Column 19, but for Channel 3.

\item[Column 34, ch3 Flag:] Same as Column 20, but for Channel 3.

\item[Column 35, ch3 Cov.:] Same as Column 21, but for Channel 3.

\item[Column 36, ch3 $M/N$:] Same as Column 22, but for Channel 3.

\item[Column 37, ch4 ID:] Same as Column 16, but for Channel 4.

\item[{Column 38, [8.0]:}] Same as Column 17, but for Channel 4.

\item[{Column 39, [8.0] unc.:}] Same as Column 18, but for Channel 4.

\item[{Column 40, [8.0] SNR:}] Same as Column 19, but for Channel 4

\item[Column 41, ch4 Flag:] Same as Column 20, but for Channel 4.

\item[Column 42, ch4 Cov.:] Same as Column 21, but for Channel 4.

\item[Column 43, ch4 $M/N$:] Same as Column 22, but for Channel 4.

\end{description}

There are a total of 1,065,565 sources in our final catalog; 
656,673 of those satisfy the GLIMPSE ``2+1" criterion and have a 
SNR$>$ 10; they are considered to be highly reliable sources.
We summarize other relevant statistics for our catalog in Table \ref{tab8}.

Two asteroids were found in the field of view at the time of
our observations. 
Asteroid Alikoski appears in the final catalog as sources SSTGC 0629833 and
SSTGC 0636843 (twice because it moved between Channel 3 and 4 coverage) 
and asteroid 459 Signe is in the final catalog as source SSTGC 0216539.

Areas close to the edges of our survey overlap with the GLIMPSE II 
(PI: Churchwell) observations.
We have compared the photometry of our sources with SNR$>$10 to the
photometry of the GLIMPSE II Highly Reliable Catalog.
There are 184,392, 160,387, 132,077, and 72,065 sources positionally matched
within 1\arcsec\  between our catalog and GLIMPSE's. 
The mean difference in magnitudes between our photometry
and GLIMPSE is 0.11, 0.06, 0.00, and 0.03 for Channels 1, 2, 3, and 4.
The standard deviations of the same differences are 0.15, 0.15, 0.17, and
0.18 magnitudes for Channels 1, 2, 3, and 4, respectively. 
The GLIMPSE photometry is expected to be have uncertainties less than
0.2 mags for most of its sources, according to the GLIMPSE Quality
Assurance Document, v1.0. Thus, the observed differences are therefore less
than the expected photometric uncertainties.

%DISCUSSION

\section{Magnitude Distributions \label{sec_num_den}}

\begin{figure}
\plotone{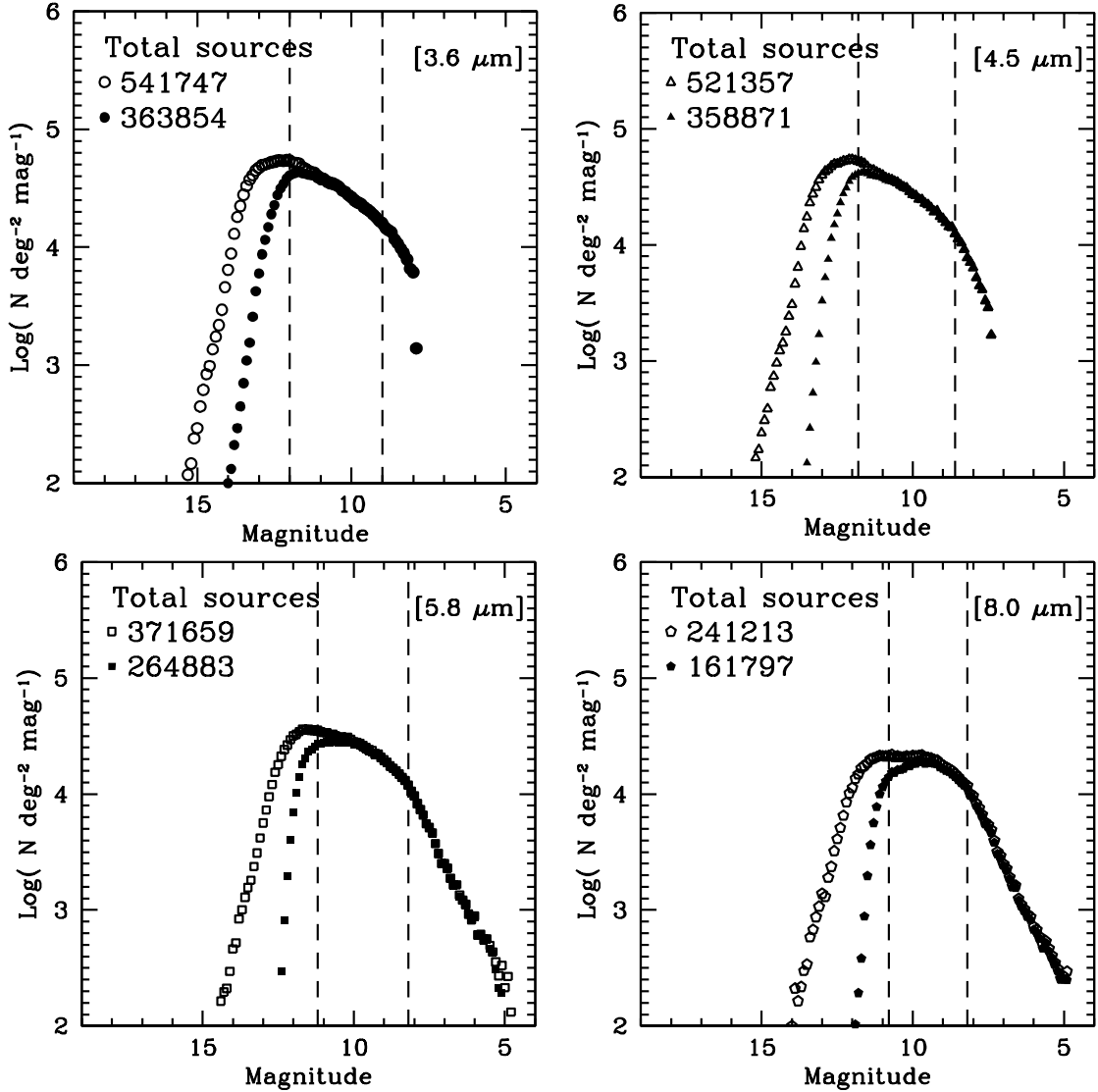}
\figcaption{Number density distribution of point sources for each
of the IRAC channels.
Only the sources located within $-1.0 \leq l \leq 1.0$ and 
$-0.7 \leq b \leq 0.7$ are included in the determination of the 
distribution.
The open symbols show the magnitude distribution for all the sources
and the filled symbols show the distribution for sources satisfying
the ``2+1" criterion and having SNR$>$10.
The total number of sources used in the determination of the 
distributions is listed on the top left side of each panel.
The bin size is 0.1 magnitudes.
The {\it dashed lines} show the limits of three brightness ranges
defined to study the distribution of point sources with $l$ and $b$.
\label{fig_num_den}}
\end{figure}

Figure \ref{fig_num_den} shows the magnitude distribution of point source
detections for each of the channels in our survey. 
Only the sources located within the uniform coverage box 
($-1.0 \leq l \leq 1.0, -0.7 \leq b \leq 0.7$) are included in 
the determination of the magnitude distributions. 
The open symbols show the magnitude distribution for all the sources 
and the filled symbols show the distribution for the sources which
satisfy the ``2+1" criterion and have a SNR greater than 10.

Figure \ref{fig_num_den} shows that the magnitude distributions 
have a similar shape in all the IRAC channels.
There is a steep slope of increasing number of stars with 
increasing magnitude at the brightest magnitudes.
This steep slope flattens around a {\it bright turnover} magnitude of 
9.0, 8.6, 8.2, and 8.2 magnitudes for Channels 1, 2, 3 and 4, respectively.
The number of stars increases more slowly with increasing magnitude
at magnitudes fainter than the {\it bright turnover}.
Finally, there is a {\it faint cutoff} followed by 
a steep slope of decreasing number of stars with increasing magnitude.
The {\it faint cutoff} for all the sources (open symbols in Figure
\ref{fig_num_den}) is 12.4, 12.1, 11.7, and 11.2 for Channels 1, 2, 
3, and 4, respectively.
The {\it faint cutoff} for the sources which satisfy the ``2+1'' 
criterion and have SNR $>$ 10 (filled symbols in Figure \ref{fig_num_den}) 
is 12.0, 11.8, 11.2, and 10.8 for Channels 1, 2, 3, and 4, respectively.
Hereafter we define the {\it bright} magnitude range as those 
magnitudes brighter than the {\it bright turnover} magnitude, 
the {\it medium} magnitude range as those magnitudes between
the {\it bright turnover} and the {\it faint cutoff}, and 
the {\it faint} magnitude range as those magnitudes fainter
than the {\it faint cutoff}.

\begin{figure}
\plotone{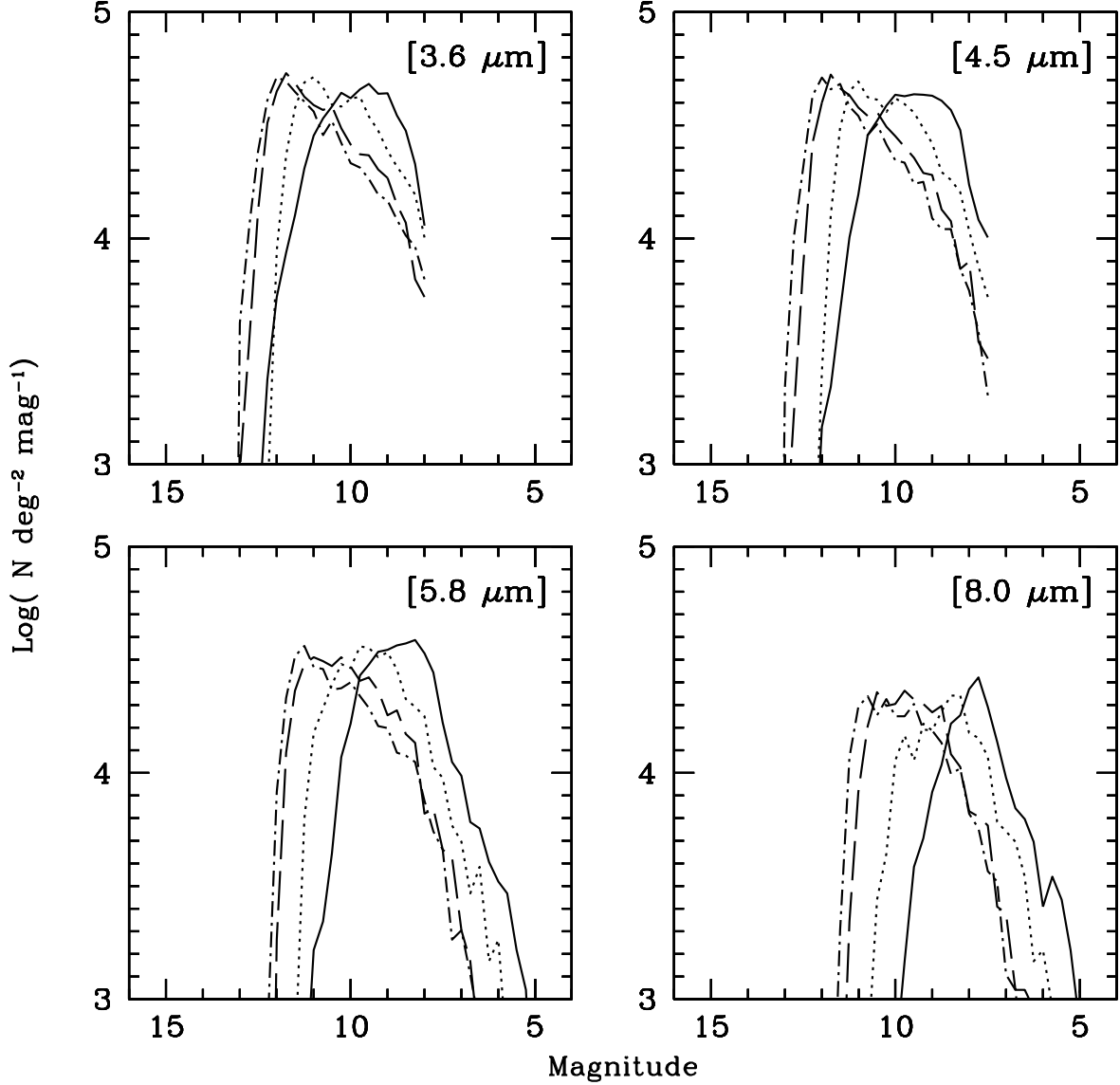}
\figcaption{Magnitude distribution within a 5\arcmin \ radius 
in locations of the survey with different source densities 
located along a diagonal going away from the GC, as plotted 
in the {\it bottom panel} of 
Figure \ref{fig_ch1}.
The bin size is 0.25 magnitudes.
The magnitude distributions of the four circular areas
are plotted  
($solid~line$: area centered on the Galactic center,
$l$=359.946, $b=-0.0378$;
$dotted~line$: area centered on $l$=0.166, $b$=0.1162;
$dashed~line$: area centered on $l$=0.386, $b$=0.2702; 
$dashed-dotted~line$: area centered on $l$=0.606, $b$=0.4242).
\label{fig_num_den_area}}
\end{figure}

We plot a subset of magnitude distributions drawn from
small regions within our GC mosaic, to understand
features of the magnitude distribution.
Only sources satisfying the ``2+1" criterion and with SNR$>$10 are 
included in the determination of these magnitude distributions.
Figure \ref{fig_num_den_area} shows the magnitude distributions 
of four circular areas located along a diagonal going away from the 
GC, but avoiding dark clouds, as plotted in the bottom panel of 
Figure \ref{fig_ch1}.
The solid, dotted, dashed and dashed-dotted lines 
show the magnitude distributions of these four
locations in the order of increasing distance to the GC.
There are about 2800, 2700, 2000, and 1200
sources in each 5\arcmin\ radius circular area
for Channels 1, 2, 3, and 4, respectively.

Figure \ref{fig_num_den_area} shows that, 
at this angular scale (10\arcmin), 
the magnitude distributions have the same shape as the 
magnitude distribution for the entire field of view 
(2.0\degree$\times$1.4\degree) as shown in Figure \ref{fig_num_den}.
The main difference among the magnitude distributions of
the circular areas is that they seem to shift towards fainter 
magnitudes with increasing distance to the GC. 
As a consequence, the circular area located at the GC 
(solid line) has at least a factor of 3 more bright sources
than the circular area located farthest from the GC 
(dashed-dotted line). For example at [3.6]=9.0, [4.5]=9.0,
[5.8]=7.0, and [8.0]=7.5 magnitudes,
the ratio of bright sources between the two areas 
is 3.0, 3.5, 5.0, and 5.5, respectively.

The {\it faint cutoff} of the magnitude distributions
can be interpreted as due to confusion.
Figure \ref{fig_num_den_area} shows that confusion is
occurring at brighter magnitudes as one gets closer to the GC.
The confusion limits suggested in Figure \ref{fig_num_den_area}
for Channels 1 and 2 range from about 8.5 mag in the circular area
located at the GC (solid line) to 13 mag in the circular area
located the farthest away from the GC (dashed-dotted line).
The same range of confusion limits for Channels 3 and 4 in
Figure \ref{fig_num_den_area} is 8.0 (solid line) to 12.0
(dashed-dotted line).

The fact that we observe more bright sources with decreasing 
distance to the GC
is consistent with previous population studies based on dereddened
$K-$band luminosity functions.
\citet{blu96} computed a dereddened $K-$band luminosity function
within 1\arcmin\ (2.3 pc) of the GC and compare it with a similar 
study at Baade's Window in the bulge.
\citet{blu96} found that both $K-$band luminosity functions had
the same slope, but there was an overabundance
of bright stars in the GC relative to Baade's Window.
\citet{nar96} constructed a dereddened $K-$band luminosity
function for a region of 16\arcmin $\times$16\arcmin\  (37$\times$37
pc) centered on the GC, but excluding the inner 2\arcmin\ of
the Galaxy.
They found a luminosity function intermediate between that
of Baade's Window and the inner 2\arcmin\ of the GC, having
an excess of luminous stars over the bulge but not as many
luminous stars as the inner 2\arcmin.
\citet{fig04} obtained dereddened 2 \micron\ luminosity 
functions, using high angular resolution observations.
They computed synthetic luminosity functions using stellar evolution
models and concluded that the observations were best fitted by models 
of continuous star formation.

All of the near infrared luminosity functions outside of the Central
Cluster show the presence of a {\it bright turnover}.
The luminosity function of the central 200 pc is known 
to have an excess of luminous stars relative to
bulge fields such as Baade's Window, and this excess 
of luminous stars increases closer to the GC
\citep{cat90,blu96,nar96,phi99,fig04}.
The variation in the number of the brightest stars 
with distance to the GC over a large area may cause 
the presence of the {\it bright turnover}.
This speculation is complicated by several effects.
First, the {\it bright turnover} for Channels 1 and 2,
at the areas close to the GC, occurs very close to the 
saturation limit (Ch. 1 and 2 magnitudes of 7.9 and 7.4, respectively).
Second, the extinction and local background in the GC are 
highly spatially variable, even within a 10\arcmin\ field of view.
Finally, stellar crowding may artificially enhance the bright
end of measured luminosity functions as pointed out by \citet{dep93}.
Higher angular resolution surveys of the GC area may indeed
improve our knowledge of the nature of the {\it bright} range sources
seen in this survey.

The magnitude distribution shown in Figure \ref{fig_num_den}
can be understood as the integral of individual magnitude
distributions such as those plotted in Figure \ref{fig_num_den_area}.
The integral of the individual magnitude distributions is
non-trivial to calculate due to the wide range in stellar density, 
extinction, background levels, and confusion within our field of view.
Complete modeling of the observed magnitude distribution is beyond
the scope of the present work and may be addressed in a
future work.

\section{Source distribution with Galactic Longitude and 
Latitude \label{sec_gcoord}}

\begin{figure}
\plotone{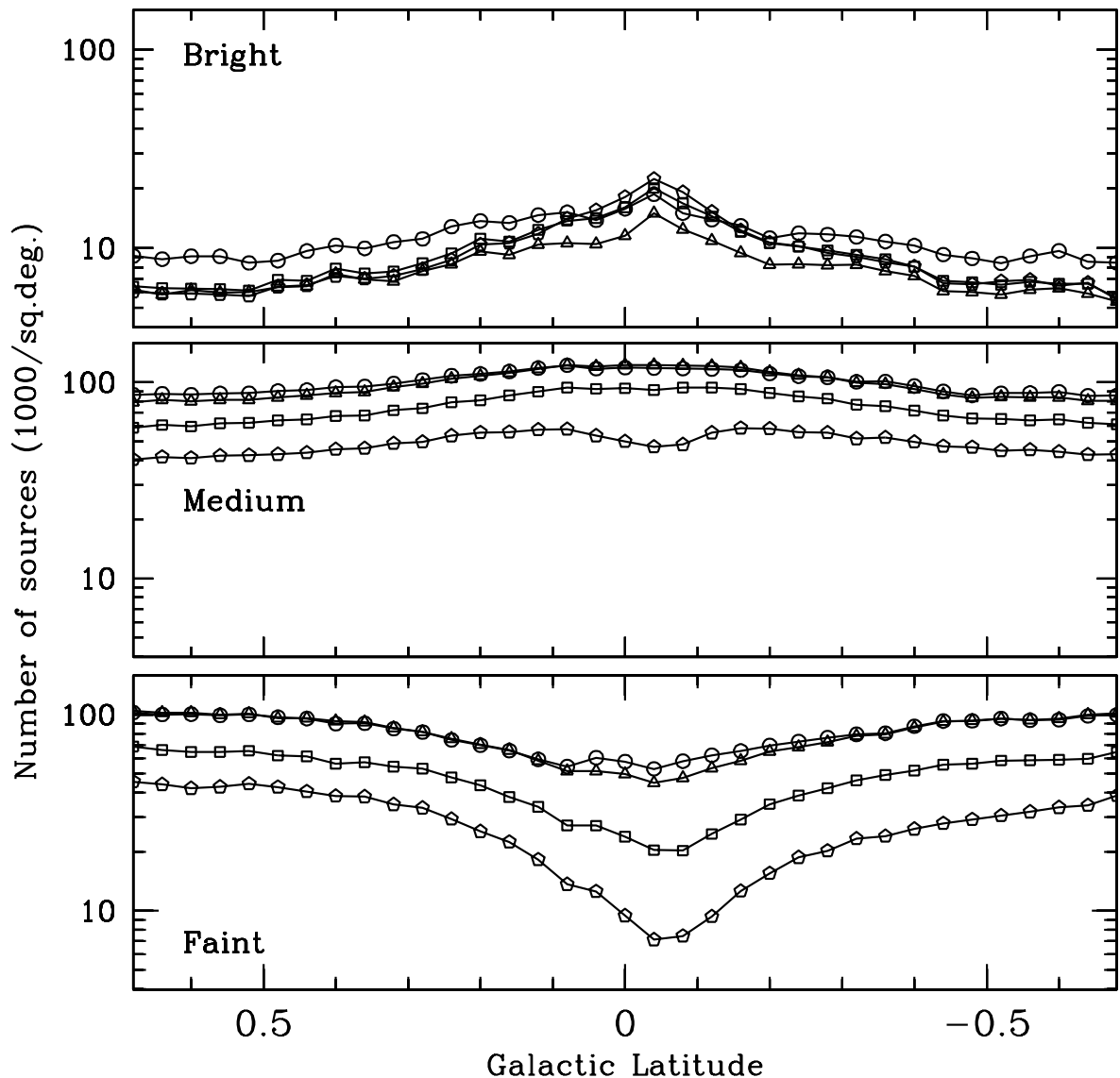}
\figcaption{Distribution of point sources with Galactic latitude
for the three brightness ranges defined in Fig. \ref{fig_num_den}.
Only the sources located within $-1.0 \leq l \leq 1.0$ and
$-0.7 \leq b \leq 0.7$ are included in the determination of the
number density distribution.
Circles, triangles, squares, and pentagons correspond to the
Galactic coordinate distributions of Channels 1, 2, 3, and 4,
respectively.
\label{fig_glat_mag}}.
\end{figure}

\begin{figure}
\plotone{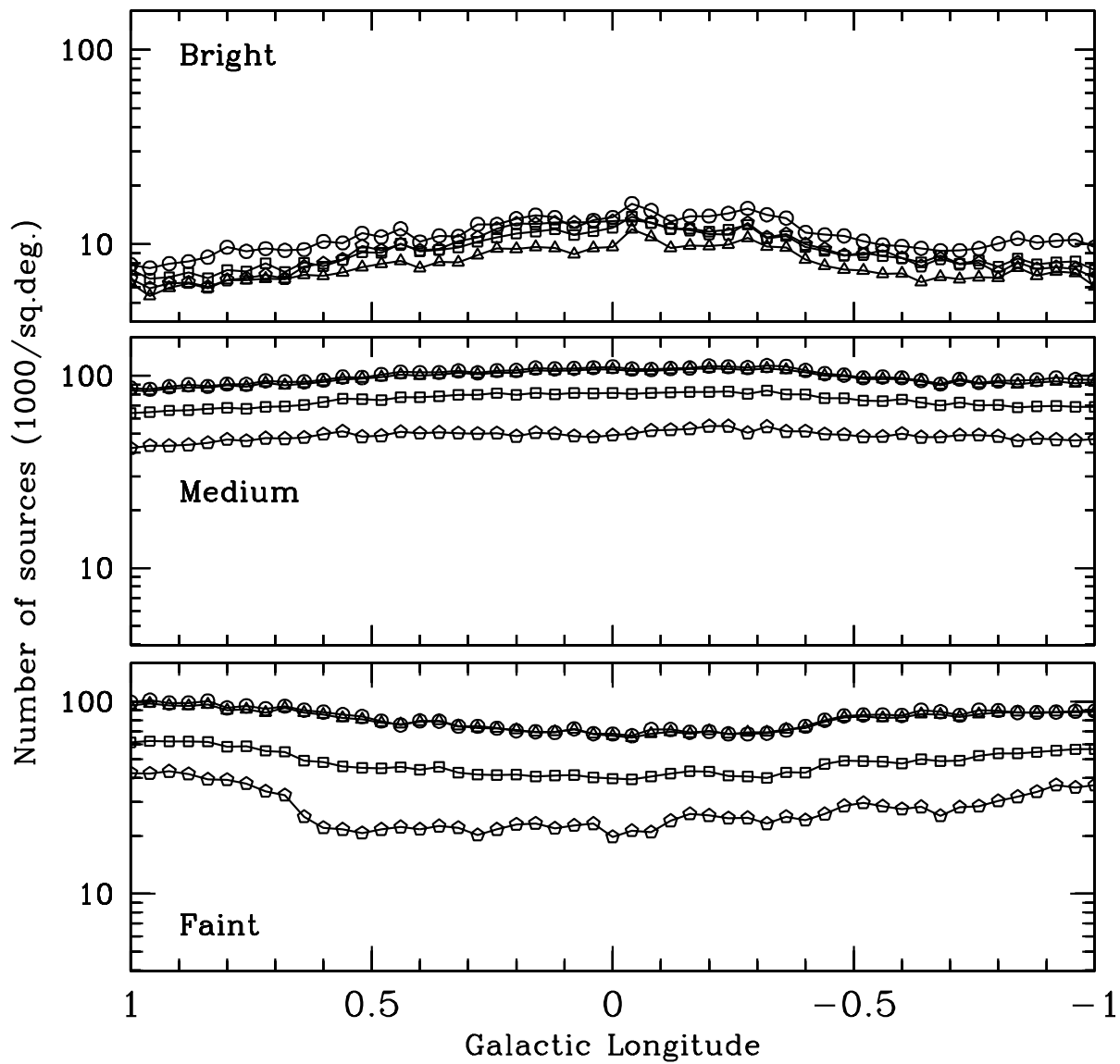}
\caption{Distribution of point sources with Galactic longitude
for the three defined  brightness ranges, as indicated.
Only the sources located within $-1.0 \leq l \leq 1.0$ and
$-0.7 \leq b \leq 0.7$ are included in the determination of the
number density distribution.
Symbols are the same as in Figure \ref{fig_glat_mag}.
\label{fig_glon_mag}}
\end{figure}

The overall density of detected point sources with 
latitude and longitude is essentially constant,
a consequence of being confusion limited.  
The relatively constant number of sources per 
circular area also indicates that our images are confusion limited.
However, interesting structure along Galactic latitude and longitude
does appear when we select sources within different magnitude ranges
(defined above).
The point source distributions along Galactic latitude and longitude 
in the different magnitude ranges are shown in Figures 
\ref{fig_glat_mag} and \ref{fig_glon_mag}.
The different panels show the distribution of the point sources
for {\it bright}, {\it medium}, and {\it faint}, as indicated.
Circles, triangles, squares, and pentagons correspond to the
Galactic coordinate distributions of Channels 1, 2, 3, and 4,
respectively.

The structure seen at the {\it bright} range is consistent with 
the fact that more {\it bright} sources are observed towards
the GC (as discussed in Section \ref{sec_num_den}).
This agrees well with previous population studies
that find an excess of luminous stars in the GC relative to
bulge fields \citep{cat90,blu96,nar96,phi99,fig04}.

The structure seen at the {\it faint} range is set by our ability
to detect faint sources. The {\it faint} range contains sources
below the lowest confusion limit for each channel, and therefore
they follow the trend of the variation of the confusion with Galactic
latitude and longitude.

The structure along Galactic latitude has a similar shape for all
the IRAC channels, but some features are more prominent at longer
wavelengths.
The sources in the {\it bright} range show a steeply decreasing
gradient with Galactic latitude, with a peak in the position of the
Galactic center. The {\it medium} range sources show a slowly
decreasing gradient with Galactic latitude.
Those sources in the {\it faint}
range show an increasing gradient with Galactic latitude, with the
minimum at the position of the Galactic center.

Figure \ref{fig_glon_mag} illustrates that
the structure along Galactic longitude also has the same shape for all
the IRAC channels.
The sources in the {\it bright} and {\it medium} brightness range 
show a slow decrease with Galactic longitude, 
with a peak at the position of the Galactic center. 
The sources in the {\it faint} range 
show an increase with Galactic longitude, with the 
minimum at the position of the Galactic center.

\section{Color-Magnitude and Color-Color Diagrams \label{sec_cmd_col}}

\begin{figure}
\epsscale{.90}
\plotone{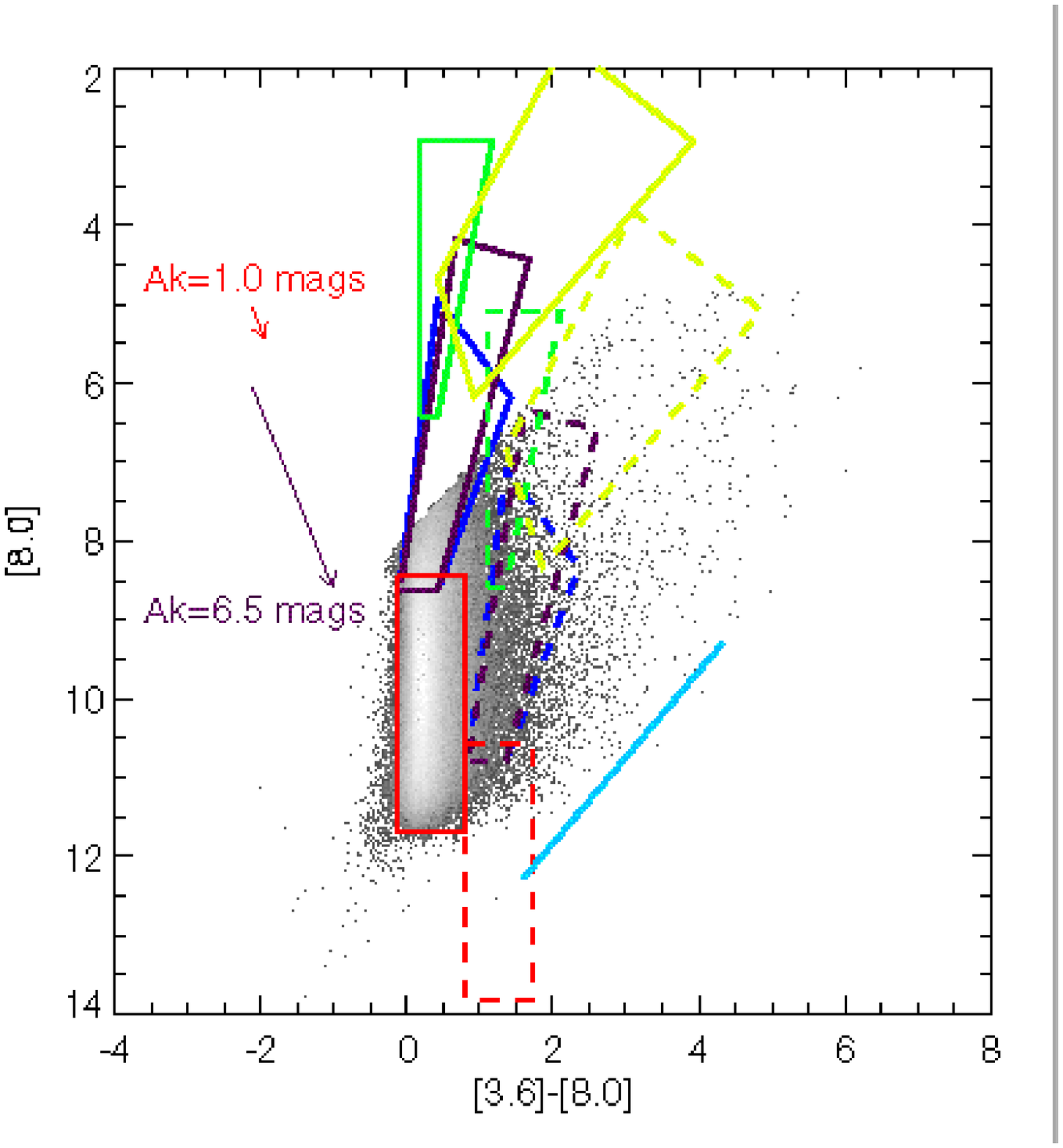}
\caption{[8.0] vs. [3.6]-[8.0] color-magnitude diagram.
The gray scale shows the number density distribution of
sources, with white being the highest density.
Only sources satisfying the ``2+1" criterion and having SNR$>$10
are plotted.
The arrows show the direction of the reddening vector, using the
extinction law from \citet{ind05}, and the minimum ($A_K$=1.0)
and maximum ($A_K$=6.5) amount of extinction measured towards the GC
\citep{blu96,sch99,dut03}.
The locations of evolved stars are taken from the 
CMD of the $Spitzer$ SAGE LMC survey \citep{blu06}
and placed at the Galactic center distance:
red giant stars ($red~boxes$), O-rich AGB stars ($blue~boxes$),
C-rich AGB stars ($purple~boxes$), extreme AGB stars ($yellow~boxes$),
and supergiant stars ($green~boxes$).
The $solid~line~boxes$ show the location of objects assuming an 
extinction of $A_K$=1.0 magnitudes, and the $dashed~line~boxes$ 
show the same boxes assuming an extinction of $A_K$=6.5
magnitudes. 
The $cyan~line$ shows the position below which background galaxies
should be located, assuming an extinction of $A_K$=1.0 magnitudes.
The location of all the point sources with colors bluer than
[3.6]-[8.0]=2.0 can be understood as evolved stars seen through
varying amounts of extinction.
\label{fig_cmd1}}
\end{figure}

\begin{figure}
\epsscale{.90}
\plotone{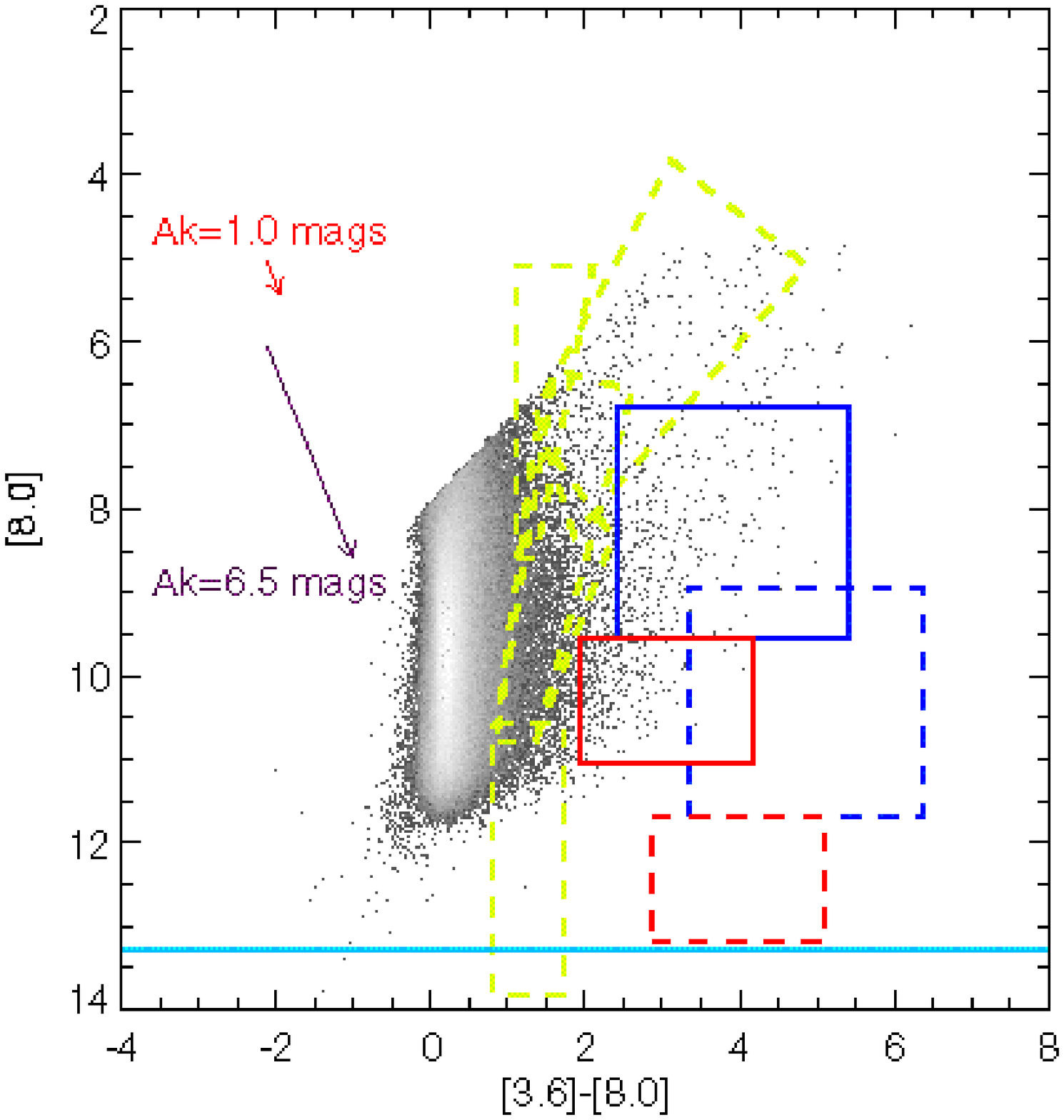}
\figcaption{[8.0] vs. [3.6]-[8.0] color-magnitude diagram.
The gray scale shows the number density distribution of
sources, with white being the highest density.
Only sources satisfying the ``2+1" criterion and having SNR$>$10
are plotted.
The arrows show the direction of the reddening vector, using the
extinction law from \citet{ind05}, and the minimum ($A_K$=1.0)
and maximum ($A_K$=6.5) amount of extinction measured towards the GC
\citep{blu96,sch99,dut03}.
The $cyan~line$ shows the [8.0] magnitude of the brightest low-mass
YSO observed in Taurus \citep{har05} at Galactic center distance.
The $red~boxes$ denote the location of 3.8 \msol \ YSOs, and the
$blue~boxes$ denote the location of 5.9 \msol \ YSOs,
as observed in the giant HII region RCW 49 by \citet{whi04}
and placed at the distance of the Galactic center.
The dotted $yellow~boxes$ show the location of evolved stars,
assuming an extinction of $A_K$=6.5 magnitudes,
as shown in Figure \ref{fig_cmd1}, for reference.
\label{fig_cmd2}}
\end{figure}

Figures \ref{fig_cmd1} and \ref{fig_cmd2} show the 
[8.0] vs. [3.6]-[8.0] color-magnitude diagram (CMD) of all the 
sources satisfying the ``2+1" criterion and having both [3.6] 
and [8.0] magnitudes with a SNR greater than 10. 
The gray scale shows the number density distribution of
sources, with white being the highest density.
The arrows show the direction of the reddening vector, using the 
extinction law from \citet{ind05}.
The red arrow shows the amount of extinction for $A_K$=1.0,
while the purple arrow shows the amount of extinction for $A_K$=6.5.
According to the extinction maps of \citet{sch99} and \citet{dut03} an
extinction of $A_K$=1.0 is observed at the edges of our survey
which we adopt as the minimum foreground extinction towards 
GC stars in our field of view.
An extinction of $A_K$=6.5 was measured as the maximum observed
extinction within 2\arcmin\  of the Galaxy by the
the near infrared photometric work of \citet{blu96}.
The highest density of points in the CMD shows a well defined sequence 
of constant [3.6]-[8.0] color, at a color of $\sim$0.2 mag. 
The distribution is skewed towards red colors, which is consistent
with varying amounts of extinction.

To determine what types of objects are seen in our survey, we have 
overplotted the location of evolved stars in the CMD of Figure 
\ref{fig_cmd1}. 
The location of evolved stars is taken from the 
[8.0] vs. [3.6]-[8.0] CMD of stars in 
the $Spitzer$ SAGE survey of the Large Magellanic
Cloud \citep{blu06}.
They determined the location of the tip of the red giant branch,
the O-rich and C-rich AGB stars, supergiant stars, and extreme
AGB stars 
from the DENIS and 2MASS analysis of LMC stars by \citet{cio06}. 
We assume a distance modulus to the LMC of 18.48 magnitudes
\citep{bor04}, and a distance to the Galactic center of 8.0 kpc
\citep{rei93} to determine their location in our observed CMD.
In Figure \ref{fig_cmd1}, the solid line boxes show the location
of objects assuming an extinction of $A_K$=1.0 magnitudes, and the dashed
line boxes show the same boxes assuming an extinction of $A_K$=6.5
magnitudes.

\citet{blu06} also noted the location of background galaxies
in their CMD.
The cyan line in Figure \ref{fig_cmd1} shows the limit
below which background galaxies should be observed,
assuming an extinction of $A_K$=1.0 magnitudes.
This line is at the edge of our observing limit, and it
would be even lower if more extinction is added to it.
We conclude that our survey is very unlikely to include
background galaxies.

In Figure \ref{fig_cmd1}, 
different colored boxes illustrate the location of
different types of stars, including
red giants (red),
O-rich stars (blue), 
C-rich stars (purple), 
extreme AGB stars (yellow), 
and supergiants (green).
The bottom of the solid red giant box (at [8.0]=11.67 mag.) 
marks the [8.0] magnitude of a K0 III star located at the GC 
observed through $A_K$=1.0 magnitudes of extinction.
There are 183,857 points sources plotted in figure \ref{fig_cmd1}.
About 78\% of the point sources shown (143,039 in total)
lie within the limits of the red solid box denoting the location 
of the red giant stars with spectral types later than K0 III.
The location of all the point sources with colors bluer than
[3.6]-[8.0]=2.0 and with [8.0] magnitudes brighter than 8.0
can be understood as evolved stars seen through
varying amounts of extinction with the range of values discussed above.

There are 917 sources in our catalog with [3.6]-[8.0] $\geq$ 2.0
and [8.0] $\geq$ 8.0.
To explore the possibility of finding young stellar objects 
among these 917 red objects, we have overplotted the location of 
YSOs in the CMD of Figure \ref{fig_cmd2}.
As in Figure \ref{fig_cmd1}, the solid line boxes show the location
of objects assuming an extinction of $A_K$=1.0 magnitudes, and the dashed
line boxes show the same boxes assuming an extinction of $A_K$=6.5
magnitudes.
The cyan line shows the [8.0] magnitude of the brightest 
low-mass
YSO observed in Taurus \citep{har05}, assuming a distance to 
Taurus of 140 pc.
This line demonstrate that low-mass YSOs cannot be detected in our survey.
\citet{whi04} studied the giant HII region RCW 49, as part of
the GLIMPSE legacy program. 
They determined the location of 2.5, 3.8, and 5.9 \msol \ YSOs, 
using the radiative transfer models from \citet{whi03}.
We assume a distance of 4.2 kpc to RCW 49 \citep{chu04} 
and a Galactic center distance of 8 kpc \citep{rei93} to 
determine their location in our observed CMD.
The red boxes denote the location of the 3.8 \msol \ YSOs, and the
blue boxes denote the location of 5.9 \msol \ YSOs.
The dotted yellow boxes show the location of evolved stars,
as reference, assuming an extinction of $A_K$=6.5 magnitudes, 
as shown in Figure \ref{fig_cmd1}.

\begin{figure}
\epsscale{.90}
\plotone{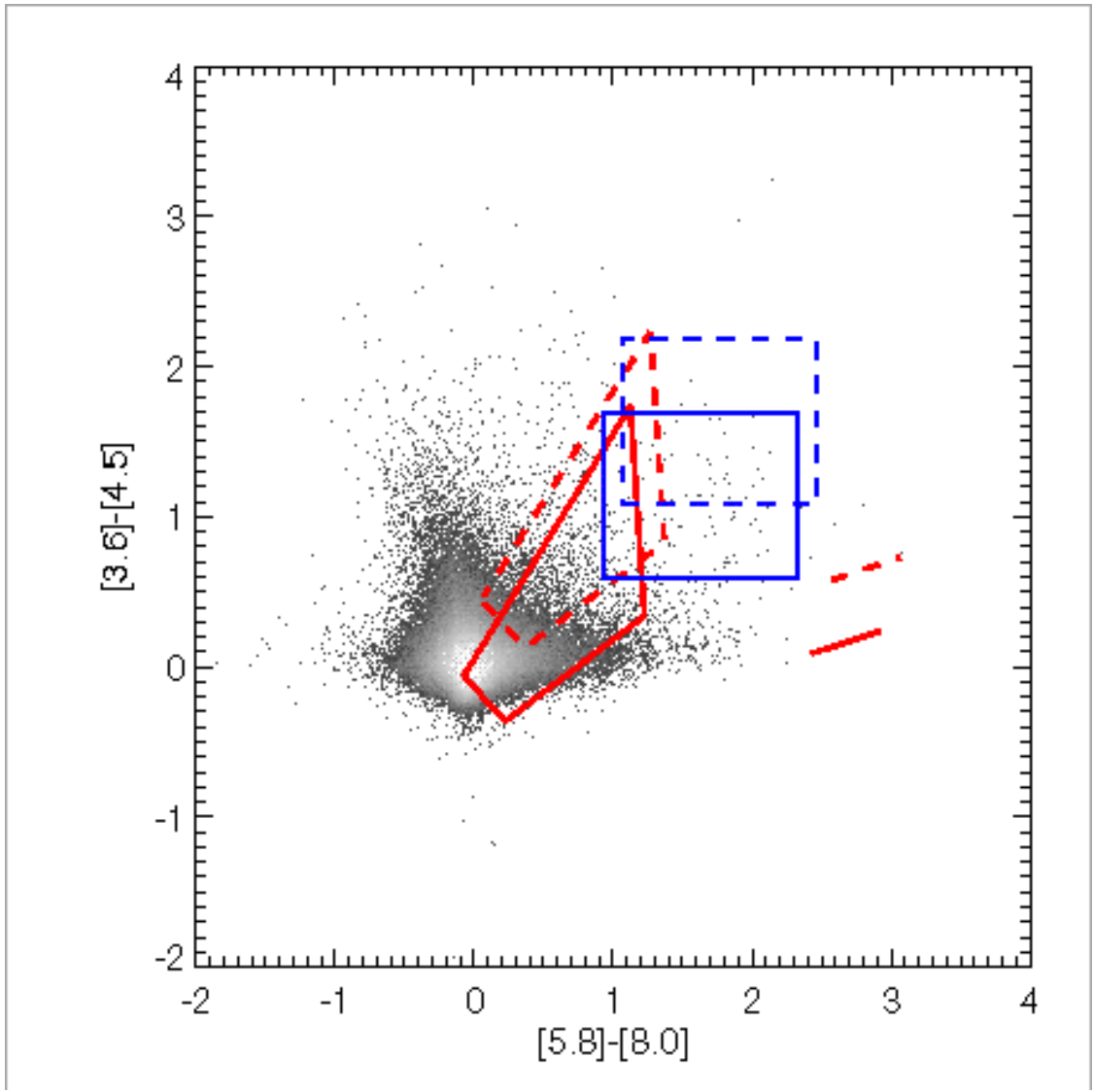}
\caption{[3.6]-[4.5] vs. [5.8]-[8.0] color-color diagram.
The gray scale shows the number density distribution of
sources, with white being the highest.
Only sources satisfying the ``2+1" criterion and having SNR$>$10 
are plotted.  
The $solid~line~boxes$ show the location
of objects assuming an extinction of $A_K$=1.0 magnitudes, 
and the $dashed~line~boxes$ show the same boxes assuming an 
extinction of $A_K$=6.5 magnitudes.
The $red~boxes$ show the average location of AGB star colors from
\citet{mar06}.
\citet{mar06} also determine the location of an AGB star 
with a thick envelope (V354 Lac), which is shown with the
$red~lines$.
The $blue~boxes$ show the location of 3.8 \msol \ and 5.9 \msol \
YSOs \citep{whi04}.
\label{fig_colcol}}
\end{figure}

Figure \ref{fig_colcol} shows the [3.6]-[4.5] vs. [5.8]-[8.0]
color-color diagram.
The gray scale shows the number density distribution of
sources, with white being the highest density.
In Figure \ref{fig_colcol}, the solid line boxes show the location
of objects assuming an extinction of $A_K$=1.0 magnitudes, and the dashed
line boxes show the same boxes assuming an extinction of $A_K$=6.5
magnitudes.
\citet{mar06} derived colors of AGB stars by convolving
observed ISO spectra with IRAC bandpasses.
The location of these derived IRAC colors are coincident with the
models of \citet{gro06} computed using stellar atmosphere
models with dust envelopes of different composition.
The red boxes show the location of AGB star colors from \citet{mar06}.
\citet{mar06} also show that an AGB star with a thick envelope
(V354 Lac) may have a unreddened [5.8]-[8.0] color between 2.4 and 2.9,
and a unreddened [3.6]-[4.5] color between 0.05 and 0.15.
The red line shows the location of V354 Lac with the corresponding
amounts of extinction. 
The blue boxes show the location of 3.8 \msol \ and 5.9 \msol \ 
YSOs \citep{whi04}.
If we consider the typical uncertainties discussed in Section \ref{full_extr},
the typical uncertainty in the [3.6]-[4.5] color is 0.06 magnitudes and
the typical uncertainty in the [5.8]-[8.0] color is 0.42 magnitudes.
There are 176,724 points sources plotted in Figure \ref{fig_colcol}.
About 38\% of the point sources shown (66,988 in total)
have zero IRAC colors within the typical uncertainties.
These point sources have been exposed to little reddening, and hence
they may be foreground objects or objects away from the Galactic plane.

As the GC is a known region of recent star formation, the possibility 
of observing a YSO population is an exciting prospect. 
We are likely to be sensitive only to the most massive YSOs (if present),
at the distance of the GC, due to confusion. 
Our survey, however, contains sources observed at different distances 
and over varying amounts of extinction.
We must carefully examine any candidate YSO population that we identify
based on IRAC colors, to distinguish foreground YSOs (in the star forming
arms along the line of sight, for instance) with a range
of masses from massive YSOs at the GC.
In addition, thick envelope AGB stars and YSOs have similar IRAC 
colors, and distinguishing one from the other will require additional 
diagnostics. 
Future work will include incorporating photometry at longer infrared 
wavelengths (e.g. ISOGAL, MSX and eventually MIPS 24 $\mu$m) and 
spectroscopy in order to best determine the nature of this population of 
objects with red IRAC colors.

%CONCLUSIONS
\section{Conclusions}

Our conclusions can be summarized as follows:

\begin{itemize}

\item A point source catalog of 1,065,565 objects is presented.
The catalog includes positions, $J$, $H$, $K_s$, [3.6], [4.5],
[5.8], and [8.0] magnitudes, and a series of flags that
assess the quality of the measurements.

\item The point source catalog is confusion limited.
The confusion limits vary by 2 to 3 magnitudes
within the field of view.
Nevertheless, the average confusion limits  
are 12.4, 12.1, 11.7, and 11.2 magnitudes
for Channels 1, 2, 3, and 4, respectively.

\item The overall distribution of point sources with Galactic
latitude and longitude is essentially constant 
(a consequence of being confusion limited), but structure
does appear when sources of different magnitude ranges 
are selected.
Bright stars show a slow decrease in number density with Galactic
longitude, and a steeper decrease with Galactic
latitude, with a peak at the position of the Galactic center.

\item Most of the point sources in our
catalog have IRAC magnitudes and colors characteristic
of red giant stars and AGB stars.
There are several hundreds of extremely red objects, however,
some of which may be massive YSOs.
Follow up observations are needed to determine the nature
of the extremely red objects.

\end{itemize}

%ACKNOWLEDGEMENTS
\acknowledgments
This work is based on observations made with the 
{\it Spitzer Space Telescope}, 
which is operated by the Jet Propulsion Laboratory, 
California Institute of Technology under a contract with NASA. 
Support for this work was provided by NASA through an 
award issued by JPL/Caltech.
This publication makes use of data products from the Two Micron 
All Sky Survey, which is a joint project of the University of 
Massachusetts and the Infrared Processing and Analysis 
Center/California Institute of Technology, funded by the National 
Aeronautics and Space Administration and the National Science Foundation.
The research described in this paper was partially carried out at the
Jet Propulsion Laboratory, California Institute of Technology, under
contract with the National Aeronautics and Space Administration.
KS thanks the NASA Faculty Fellowship Program for financial
support and the hospitality of JPL's Long Wavelength Center
and the Spitzer Science Center.
We thank S. Carey, P. Lowrance, R. Blum, C. Koresko, D. Shupe, M. Meade, 
and B. Babler for enlightening discussions.

%REFERENCES
%BIBLIOGRAPHY

%TABLES

\clearpage
\input{tab1}
\input{stub.tab2}
\input{stub.tab3}
\input{stub.tab4}
\input{stub.tab5}
\input{stub.tab6}
\input{stub.tab7}
\input{tab8}

%FIGURES

\clearpage

\clearpage

\clearpage

\clearpage

\clearpage

\clearpage

\clearpage

\clearpage

\clearpage

\clearpage

\clearpage

\clearpage

\clearpage

\end{document}

%% file: tab1.tex
%\documentclass[12pt,preprint]{aastex}
%\begin{document}

\begin{deluxetable}{cccccccc}
\tabletypesize{\scriptsize}
%\rotate
\tablecolumns{7}
\tablenum{1}
\tablecaption{Mean IRAC colors of foreground sources.
\label{tab1}}
\tablewidth{0pc}
\tablehead{
\colhead{} & \colhead{Color} & 
\colhead{Mean} & \colhead{Std. Dev.} & 
\colhead{Mean} & \colhead{Std. Dev.} & 
\colhead{Mean} & \colhead{Std. Dev.} \\
\colhead{} & \colhead{} &
\colhead{PRF}      & \colhead{PRF} &
\colhead{Aperture} & \colhead{Aperture} &
\colhead{Final}    & \colhead{Final} }
\startdata
 & [3.6]-[4.5] & 0.02 & 0.15 & 0.02 & 0.13 & 0.01 & 0.15  \\
 & [3.6]-[5.8] & 0.23 & 0.25 & 0.08 & 0.26 & 0.11 & 0.25  \\
 & [3.6]-[8.0] & 0.37 & 0.42 & 0.10 & 0.45 & 0.16 & 0.42  \\
 & [4.5]-[5.8] & 0.22 & 0.18 & 0.06 & 0.19 & 0.10 & 0.18  \\
 & [4.5]-[8.0] & 0.35 & 0.38 & 0.08 & 0.41 & 0.15 & 0.38  \\
 & [5.8]-[8.0] & 0.13 & 0.34 & 0.02 & 0.36 & 0.05 & 0.34  \\
\enddata
\end{deluxetable}

%\end{document}

%% file: stub.tab2.tex
\begin{deluxetable}{rcccccccccc}
\tabletypesize{\scriptsize}
%\rotate
\tablecolumns{11}
\tablecaption{Galactic Center IRAC 1 Source Lists\label{tab2}}
\tablewidth{0pc}
\tablenum{2}
\tablehead{
\colhead{Source ID} & 
\colhead{Channel} &
\colhead{R.A.} &
\colhead{Dec.} &
\colhead{$l$} &
\colhead{$b$} &
\colhead{Flux} &
\colhead{Flux Unc.} &
\colhead{$N_{obs}$} &
\colhead{S/N} &
\colhead{Method} \\ 
\colhead{} & 
\colhead{} &
\colhead{(J2000)} &
\colhead{(J2000)} &
\colhead{} &
\colhead{} &
\colhead{(mJy)} &
\colhead{(mJy)} &
\colhead{} &
\colhead{} &
\colhead{} }
\startdata
GC-IRAC1-000001 & 1 & 17 40 11.07 & -29 20 57.3 & 359.02435920 & 0.79220897 &    13.590 &     0.372 & 2 & 815.0 & APC      \\ 
GC-IRAC1-000002 & 1 & 17 40 11.15 & -29 20 52.3 & 359.02569293 & 0.79268906 &     8.159 &     0.135 & 2 & 321.7 & APC      \\ 
GC-IRAC1-000003 & 1 & 17 40 11.19 & -29 21 04.8 & 359.02283701 & 0.79073782 &     5.534 &     0.342 & 2 & 11.4 & PRF      \\ 
GC-IRAC1-000004 & 1 & 17 40 11.20 & -29 21 14.0 & 359.02068282 & 0.78935613 &     2.645 &     0.116 & 2 & 5.7 & PRF      \\ 
GC-IRAC1-000005 & 1 & 17 40 11.70 & -29 21 15.5 & 359.02129310 & 0.78760658 &     7.707 &     0.166 & 2 & 31.6 & PRF      \\ 
GC-IRAC1-000006 & 1 & 17 40 11.88 & -29 21 04.3 & 359.02428009 & 0.78869809 &     2.875 &     0.103 & 2 & 16.5 & PRF      \\ 
GC-IRAC1-000007 & 1 & 17 40 11.89 & -29 20 57.9 & 359.02581916 & 0.78959725 &     2.958 &     0.109 & 2 & 17.0 & PRF      \\ 
GC-IRAC1-000008 & 1 & 17 40 11.97 & -29 16 23.4 & 359.09058043 & 0.82984460 &     1.704 &     0.081 & 2 & 12.9 & PRF      \\ 
GC-IRAC1-000009 & 1 & 17 40 12.26 & -29 21 03.8 & 359.02512173 & 0.78758333 &     1.661 &     0.082 & 2 & 12.0 & PRF      \\ 
GC-IRAC1-000010 & 1 & 17 40 12.30 & -29 16 14.5 & 359.09331294 & 0.83014414 &    13.470 &     0.227 & 2 & 86.2 & PRF      \\ 
\enddata 
\tablecomments{Table \ref{tab2} is published in its entirety 
in the electronic edition of the {\it Astrophysical Journal}.
A portion is shown here for guidance regarding its form and content.}
\end{deluxetable}

%% file: stub.tab3.tex
\begin{deluxetable}{rcccccccccc}
\tabletypesize{\scriptsize}
%\rotate
\tablecolumns{11}
\tablecaption{Galactic Center IRAC 2 Source Lists\label{tab3}}
\tablewidth{0pc}
\tablenum{3}
\tablehead{
\colhead{Source ID} & 
\colhead{Channel} &
\colhead{R.A.} &
\colhead{Dec.} &
\colhead{$l$} &
\colhead{$b$} &
\colhead{Flux} &
\colhead{Flux Unc.} &
\colhead{$N_{obs}$} &
\colhead{S/N} &
\colhead{Method} \\ 
\colhead{} & 
\colhead{} &
\colhead{(J2000)} &
\colhead{(J2000)} &
\colhead{} &
\colhead{} &
\colhead{(mJy)} &
\colhead{(mJy)} &
\colhead{} &
\colhead{} &
\colhead{} }
\startdata
GC-IRAC2-000001 & 2 & 17 40 11.03 & -29 28 01.2 & 358.92449479 & 0.72982023 &    12.768 &     0.554 & 2 & 124.3 & PRF      \\ 
GC-IRAC2-000002 & 2 & 17 40 11.48 & -29 27 52.0 & 358.92753029 & 0.72979809 &     1.755 &     0.074 & 2 & 19.5 & PRF      \\ 
GC-IRAC2-000003 & 2 & 17 40 11.48 & -29 28 02.5 & 358.92505842 & 0.72822659 &     2.197 &     0.086 & 2 & 21.4 & PRF      \\ 
GC-IRAC2-000004 & 2 & 17 40 11.81 & -29 27 56.4 & 358.92712847 & 0.72814594 &     1.282 &     0.070 & 2 & 12.5 & PRF      \\ 
GC-IRAC2-000005 & 2 & 17 40 12.19 & -29 23 09.4 & 358.99542952 & 0.76930396 &    20.899 &     0.290 & 2 & 93.5 & PRF      \\ 
GC-IRAC2-000006 & 2 & 17 40 12.21 & -29 27 46.2 & 358.93030479 & 0.72838905 &     8.130 &     0.155 & 2 & 68.4 & PRF      \\ 
GC-IRAC2-000007 & 2 & 17 40 12.33 & -29 22 59.1 & 358.99812850 & 0.77039026 &     1.567 &     0.077 & 2 & 10.0 & PRF      \\ 
GC-IRAC2-000008 & 2 & 17 40 12.58 & -29 27 53.1 & 358.92940792 & 0.72623917 &    69.317 &     0.679 & 2 & 411.6 & PRF      \\ 
GC-IRAC2-000009 & 2 & 17 40 12.67 & -29 23 06.4 & 358.99705877 & 0.76823969 &    41.474 &     0.482 & 2 & 236.6 & PRF      \\ 
GC-IRAC2-000010 & 2 & 17 40 12.78 & -29 27 48.3 & 358.93092495 & 0.72633266 &     2.025 &     0.092 & 2 & 10.2 & PRF      \\ 
\enddata 
\tablecomments{Table \ref{tab3} is published in its entirety 
in the electronic edition of the {\it Astrophysical Journal}.
A portion is shown here for guidance regarding its form and content.}
\end{deluxetable}

%% file: stub.tab4.tex
\begin{deluxetable}{rcccccccccc}
\tabletypesize{\scriptsize}
%\rotate
\tablecolumns{11}
\tablecaption{Galactic Center IRAC 3 Source Lists\label{tab4}}
\tablewidth{0pc}
\tablenum{4}
\tablehead{
\colhead{Source ID} & 
\colhead{Channel} &
\colhead{R.A.} &
\colhead{Dec.} &
\colhead{$l$} &
\colhead{$b$} &
\colhead{Flux} &
\colhead{Flux Unc.} &
\colhead{$N_{obs}$} &
\colhead{S/N} &
\colhead{Method} \\ 
\colhead{} & 
\colhead{} &
\colhead{(J2000)} &
\colhead{(J2000)} &
\colhead{} &
\colhead{} &
\colhead{(mJy)} &
\colhead{(mJy)} &
\colhead{} &
\colhead{} &
\colhead{} }
\startdata
GC-IRAC3-000001 & 3 & 17 40 11.66 & -29 21 22.8 & 359.01949975 & 0.78664857 &    36.690 &     0.678 & 2 & 96.3 & PRF      \\ 
GC-IRAC3-000002 & 3 & 17 40 11.72 & -29 21 15.4 & 359.02136276 & 0.78755549 &     3.452 &     0.254 & 2 & 15.3 & PRF      \\ 
GC-IRAC3-000003 & 3 & 17 40 12.31 & -29 16 14.5 & 359.09332451 & 0.83010156 &     6.246 &     0.309 & 2 & 36.0 & PRF      \\ 
GC-IRAC3-000004 & 3 & 17 40 12.56 & -29 21 15.5 & 359.02295870 & 0.78495848 &     8.214 &     0.324 & 2 & 38.2 & PRF      \\ 
GC-IRAC3-000005 & 3 & 17 40 12.67 & -29 21 19.8 & 359.02215756 & 0.78396611 &     1.979 &     0.230 & 2 & 8.3 & PRF      \\ 
GC-IRAC3-000006 & 3 & 17 40 13.09 & -29 16 34.1 & 359.09022025 & 0.82480989 &     2.374 &     0.229 & 2 & 12.2 & PRF      \\ 
GC-IRAC3-000007 & 3 & 17 40 13.17 & -29 16 23.8 & 359.09280942 & 0.82608296 &     9.377 &     0.337 & 2 & 47.5 & PRF      \\ 
GC-IRAC3-000008 & 3 & 17 40 13.24 & -29 16 40.2 & 359.08907321 & 0.82344698 &    29.050 &     1.770 & 2 & 149.5 & PRF      \\ 
GC-IRAC3-000009 & 3 & 17 40 13.30 & -29 21 25.5 & 359.02202594 & 0.78121008 &    14.530 &     0.401 & 2 & 61.4 & PRF      \\ 
GC-IRAC3-000010 & 3 & 17 40 13.45 & -29 25 34.0 & 358.96380980 & 0.74408637 &    41.970 &     2.710 & 2 & 105.3 & PRF      \\ 
\enddata 
\tablecomments{Table \ref{tab4} is published in its entirety 
in the electronic edition of the {\it Astrophysical Journal}.
A portion is shown here for guidance regarding its form and content.}
\end{deluxetable}

%% file: stub.tab5.tex
\begin{deluxetable}{rcccccccccc}
\tabletypesize{\scriptsize}
%\rotate
\tablecolumns{11}
\tablecaption{Galactic Center IRAC 4 Source Lists\label{tab5}}
\tablewidth{0pc}
\tablenum{5}
\tablehead{
\colhead{Source ID} & 
\colhead{Channel} &
\colhead{R.A.} &
\colhead{Dec.} &
\colhead{$l$} &
\colhead{$b$} &
\colhead{Flux} &
\colhead{Flux Unc.} &
\colhead{$N_{obs}$} &
\colhead{S/N} &
\colhead{Method} \\ 
\colhead{} & 
\colhead{} &
\colhead{(J2000)} &
\colhead{(J2000)} &
\colhead{} &
\colhead{} &
\colhead{(mJy)} &
\colhead{(mJy)} &
\colhead{} &
\colhead{} &
\colhead{} }
\startdata
GC-IRAC4-000001 & 4 & 17 40 12.18 & -29 32 20.1 & 358.86576803 & 0.68811768 &     3.358 &     0.261 & 2 & 49000.0 & APC      \\ 
GC-IRAC4-000002 & 4 & 17 40 12.23 & -29 27 46.2 & 358.93034169 & 0.72836161 &     2.477 &     0.563 & 1 & 10.2 & PRF      \\ 
GC-IRAC4-000003 & 4 & 17 40 12.26 & -29 23 20.6 & 358.99291905 & 0.76741035 &     4.660 &     0.489 & 2 & 20.0 & PRF      \\ 
GC-IRAC4-000004 & 4 & 17 40 12.58 & -29 23 17.3 & 358.99431916 & 0.76693178 &     7.815 &     0.424 & 2 & 33.6 & PRF      \\ 
GC-IRAC4-000005 & 4 & 17 40 12.59 & -29 27 53.1 & 358.92940149 & 0.72622137 &    24.106 &     0.649 & 2 & 77.2 & PRF      \\ 
GC-IRAC4-000006 & 4 & 17 40 12.68 & -29 23 06.4 & 358.99707024 & 0.76821160 &    17.530 &     0.551 & 2 & 76.8 & PRF      \\ 
GC-IRAC4-000007 & 4 & 17 40 12.98 & -29 23 16.1 & 358.99537641 & 0.76586405 &     1.402 &     0.336 & 2 & 5.1 & PRF      \\ 
GC-IRAC4-000008 & 4 & 17 40 12.99 & -29 32 43.3 & 358.86187264 & 0.68219639 &    33.694 &     5.370 & 2 & 70.9 & PRF      \\ 
GC-IRAC4-000009 & 4 & 17 40 13.14 & -29 23 19.0 & 358.99500161 & 0.76494954 &     2.809 &     0.378 & 2 & 13.2 & PRF      \\ 
GC-IRAC4-000010 & 4 & 17 40 13.28 & -29 27 58.8 & 358.92940310 & 0.72326540 &     6.769 &     0.448 & 2 & 29.1 & PRF      \\ 
\enddata 
\tablecomments{Table \ref{tab5} is published in its entirety 
in the electronic edition of the {\it Astrophysical Journal}.
A portion is shown here for guidance regarding its form and content.}
\end{deluxetable}

%% file: stub.tab6.tex
\begin{deluxetable}{rlrrrrrrrrrrrr}
\tabletypesize{\scriptsize}
\rotate
\tablecolumns{14}
\tablecaption{Galactic Center IRAC sub-array photometry\label{tab6}}
\tablewidth{0pc}
\tablenum{6}
\tablehead{
\colhead{ID} &
\colhead{Field} &
\colhead{R.A.} &
\colhead{Dec.} &
\colhead{$l$} &
\colhead{$b$} &
\colhead{[3.6] mag.} &
\colhead{[3.6] Unc.} &
\colhead{[4.5] mag.} &
\colhead{[4.5] Unc.} &
\colhead{[5.8] mag.} &
\colhead{[5.8] Unc.} &
\colhead{[8.0] mag.} &
\colhead{[8.0] Unc.} \\ 
\colhead{} &
\colhead{} &
\colhead{(J2000)} &
\colhead{(J2000)} &
\colhead{} &
\colhead{} &
\colhead{(mag)} &
\colhead{(mag)} &
\colhead{(mag)} &
\colhead{(mag)} &
\colhead{(mag)} &
\colhead{(mag)} &
\colhead{(mag)} &
\colhead{(mag)} }
\startdata
1 & sat & 17 46 02.16 & -28 57 23.6 & 0.02995207 & -0.08824025 & 4.197 & 0.003 & 2.705 & 0.002 & 1.610 & 0.001 & 0.970 & 0.001 \\ 
2 & sat & 17 44 51.28 & -29 24 55.0 & 359.50406846 & -0.10735842 & 3.078 & 0.002 & 2.524 & 0.002 & 2.030 & 0.002 & 1.697 & 0.002 \\ 
3 & sat & 17 47 44.84 & -28 26 36.5 & 0.66327339 & -0.14292245 & 4.701 & 0.004 & 3.196 & 0.002 & 2.091 & 0.002 & 1.491 & 0.002 \\ 
4 & sat & 17 45 28.65 & -28 56 05.0 & 359.98495800 & 0.02744322 & 6.580 & 0.010 & 5.205 & 0.006 & 3.692 & 0.004 & 1.428 & 0.002 \\ 
5 & sat & 17 45 01.66 & -29 26 05.1 & 359.50714659 & -0.14965396 & 2.701 & 0.001 & 2.300 & 0.001 & 1.813 & 0.002 & 1.400 & 0.002 \\ 
6 & sat & 17 46 45.24 & -28 15 47.6 & 0.70422207 & 0.13739486 & 2.997 & 0.002 & 1.760 & 0.001 & 0.939 & 0.001 & 0.629 & 0.001 \\ 
7 & sat & 17 47 37.62 & -29 03 28.1 & 0.12387807 & -0.43814137 & 2.219 & 0.001 & 1.942 & 0.001 & 1.492 & 0.001 & 1.034 & 0.001 \\ 
8 & sat & 17 47 19.87 & -29 11 54.7 & 359.97003487 & -0.45571545 & 6.162 & 0.018 & 5.067 & 0.006 & 4.159 & 0.006 & 3.167 & 0.005 \\ 
9 & sat & 17 47 20.17 & -29 11 59.1 & 359.96954232 & -0.45730964 & 6.533 & 0.010 & 3.907 & 0.003 & 2.545 & 0.002 & 1.843 & 0.002 \\ 
10 & sat & 17 44 34.94 & -29 04 35.5 & 359.76181240 & 0.12032233 & 5.141 & 0.004 & 3.940 & 0.003 & 2.919 & 0.003 & 2.151 & 0.002 \\ 
\enddata 
\tablecomments{Table \ref{tab6} is published in its entirety 
in the electronic edition of the {\it Astrophysical Journal}.
A portion is shown here for guidance regarding its form and content.}
\end{deluxetable}

%% file: stub.tab7.tex
\begin{deluxetable}{rrrrrcccrrrrrrc}
\tabletypesize{\scriptsize}
\rotate
\tablecolumns{15}
\tablenum{7}
\tablecaption{Galactic Center 2MASS/IRAC Catalog \label{tab7}}
\tablewidth{0pc}
\tablehead{
\colhead{Source ID} & 
\colhead{R.A.} &
\colhead{Dec.} &
\colhead{$l$} &
\colhead{$b$} &
\colhead{"2+1"} &
\colhead{Pos.} &
\colhead{2MASS ID} &
\colhead{$J$} &
\colhead{$J$ unc.} &
\colhead{$H$} &
\colhead{$H$ unc.} &
\colhead{$K_s$} &
\colhead{$K_s$ unc.} &
\colhead{Qual} \\
\colhead{} & 
\colhead{(J2000)} &
\colhead{(J2000)} &
\colhead{} &
\colhead{} &
\colhead{Flag} &
\colhead{Flag} &
\colhead{} &
\colhead{(mag)} &
\colhead{(mag)} &
\colhead{(mag)} &
\colhead{(mag)} &
\colhead{(mag)} &
\colhead{(mag)} &
\colhead{Flag}  \\
\colhead{(1)} & 
\colhead{(2)} &
\colhead{(3)} &
\colhead{(4)} &
\colhead{(5)} &
\colhead{(6)} &
\colhead{(7)} &
\colhead{(8)} &
\colhead{(9)} &
\colhead{(10)} &
\colhead{(11)} &
\colhead{(12)} &
\colhead{(13)} &
\colhead{(14)} &
\colhead{(15)} }
\startdata
SSTGC 0156368 & 17 42 58.64 & -28 39 28.8 &359.93395 & 0.63899 & 1 & 1 &             none &-9.999 &-9.999 &-9.999 &-9.999 &-9.999 &-9.999 &ZZZ \\
SSTGC 0156369 & 17 42 58.64 & -28 52 13.2 &359.75333 & 0.52735 & 1 & 1 & 17425864-2852131 &13.739 &0.067 &10.857 &0.048 &9.584 &0.040 &AEA \\
SSTGC 0156370 & 17 42 58.64 & -28 53 44.9 &359.73166 & 0.51395 & 0 & 1 & 17425863-2853449 &15.413 &0.043 &13.455 &0.048 &12.628 &0.066 &DAA \\
SSTGC 0156371 & 17 42 58.64 & -29 26 51.8 &359.26219 & 0.22374 & 1 & 1 & 17425863-2926518 &15.605 &-9.999 &12.276 &0.037 &10.682 &0.054 &UAA \\
SSTGC 0156372 & 17 42 58.64 & -29 32 59.9 &359.17521 & 0.16997 & 0 & 1 &             none &-9.999 &-9.999 &-9.999 &-9.999 &-9.999 &-9.999 &ZZZ \\
\enddata 
\tablecomments{Table \ref{tab7} is published in its entirety 
in the electronic edition of the {\it Astrophysical Journal}.
A portion is shown here for guidance regarding its form and content.}
\end{deluxetable}

\begin{deluxetable}{ccrrrccccrrrccc}
\tabletypesize{\scriptsize}
\rotate
\tablecolumns{14}
\tablenum{7}
\tablecaption{Galactic Center 2MASS/IRAC Catalog (cont.)}
\tablewidth{0pc}
\tablehead{
\colhead{Source ID} &
\colhead{ch1 ID} &
\colhead{ch1 Mag} &
\colhead{ch1 unc.} &
\colhead{ch1 SNR} &
\colhead{ch1 } &
\colhead{ch1 } &
\colhead{ch1 } &
\colhead{ch2 ID} &
\colhead{ch2 Mag} &
\colhead{ch2 unc.} &
\colhead{ch2 SNR} &
\colhead{ch2 } &
\colhead{ch2 } &
\colhead{ch2 } \\ 
\colhead{} &
\colhead{} &
\colhead{(mag)} &
\colhead{(mag)} &
\colhead{} &
\colhead{Flag} &
\colhead{Cov.} &
\colhead{$M/N$} &
\colhead{} &
\colhead{(mag)} &
\colhead{(mag)} &
\colhead{} &
\colhead{Flag} &
\colhead{Cov.} &
\colhead{$M/N$} \\ 
\colhead{(1)} &
\colhead{(16)} &
\colhead{(17)} &
\colhead{(18)} &
\colhead{(19)} &
\colhead{(20)} &
\colhead{(21)} &
\colhead{(22)} &
\colhead{(23)} &
\colhead{(24)} &
\colhead{(25)} &
\colhead{(26)} &
\colhead{(27)} &
\colhead{(28)} &
\colhead{(29)} }
\startdata
SSTGC 0156368 & GC-IRAC1-109085 &  13.386 &   0.045 &       5.2 & 1 & 6 & 1.00 & GC-IRAC2-103799 &  12.799 &   0.036 &       9.2 & 1 & 5 & 1.00 \\ 
SSTGC 0156369 & GC-IRAC1-109089 &   8.606 &   0.007 &     389.9 & 1 & 5 & 1.00 & GC-IRAC2-103795 &   8.747 &   0.008 &     336.2 & 1 & 4 & 1.00 \\ 
SSTGC 0156370 & none &  -9.999 &  -9.999 &      -9.9 & 0 & 6 & 0.00 & none &  -9.999 &  -9.999 &      -9.9 & 0 & 4 & 0.00 \\ 
SSTGC 0156371 & GC-IRAC1-109087 &   9.629 &   0.010 &     104.5 & 1 & 4 & 1.00 & GC-IRAC2-103797 &   9.657 &   0.010 &      78.7 & 1 & 5 & 1.00 \\ 
SSTGC 0156372 & none &  -9.999 &  -9.999 &      -9.9 & 0 & 5 & 0.00 & none &  -9.999 &  -9.999 &      -9.9 & 0 & 6 & 0.00 \\ 
\enddata 
\tablecomments{Table \ref{tab7} is published in its entirety 
in the electronic edition of the {\it Astrophysical Journal}.
A portion is shown here for guidance regarding its form and content.}
\end{deluxetable}

\begin{deluxetable}{ccrrrccccrrrccc}
\tabletypesize{\scriptsize}
\rotate
\tablecolumns{14}
\tablenum{7}
\tablecaption{Galactic Center 2MASS/IRAC Catalog (cont.)}
\tablewidth{0pc}
\tablehead{
\colhead{Source ID} &
\colhead{ch3 ID} &
\colhead{ch3 Mag} &
\colhead{ch3 unc.} &
\colhead{ch3 SNR} &
\colhead{ch3 } &
\colhead{ch3 } &
\colhead{ch3 } &
\colhead{ch4 ID} &
\colhead{ch4 Mag} &
\colhead{ch4 unc.} &
\colhead{ch4 SNR} &
\colhead{ch4 } &
\colhead{ch4 } &
\colhead{ch4 } \\ 
\colhead{} &
\colhead{} &
\colhead{(mag)} &
\colhead{(mag)} &
\colhead{} &
\colhead{Flag} &
\colhead{Cov.} &
\colhead{$M/N$} &
\colhead{} &
\colhead{(mag)} &
\colhead{(mag)} &
\colhead{} &
\colhead{Flag} &
\colhead{Cov.} &
\colhead{$M/N$}\\ 
\colhead{(1)} &
\colhead{(30)} &
\colhead{(31)} &
\colhead{(32)} &
\colhead{(33)} &
\colhead{(34)} &
\colhead{(35)} &
\colhead{(36)} &
\colhead{(37)} &
\colhead{(38)} &
\colhead{(39)} &
\colhead{(40)} &
\colhead{(41)} &
\colhead{(42)} &
\colhead{(43)} }
\startdata
SSTGC 0156368 & none &  -9.999 &  -9.999 &      -9.9 & 0 & 6 & 0.00 & none &  -9.999 &  -9.999 &      -9.9 & 0 & 5 & 0.00 \\ 
SSTGC 0156369 & GC-IRAC3-072380 &   8.383 &   0.012 &     233.9 & 1 & 5 & 1.00 & GC-IRAC4-049713 &   8.496 &   0.024 &     121.5 & 1 & 4 & 1.00 \\ 
SSTGC 0156370 & none &  -9.999 &  -9.999 &      -9.9 & 0 & 5 & 0.00 & none &  -9.999 &  -9.999 &      -9.9 & 0 & 4 & 0.00 \\ 
SSTGC 0156371 & GC-IRAC3-072378 &   9.293 &   0.023 &      59.2 & 1 & 4 & 1.00 & GC-IRAC4-049712 &   9.387 &   0.057 &      32.5 & 1 & 5 & 1.00 \\ 
SSTGC 0156372 & none &  -9.999 &  -9.999 &      -9.9 & 0 & 5 & 0.00 & GC-IRAC4-049716 &  11.251 &   0.242 &       6.3 & 1 & 5 & 1.00 \\ 
\enddata 
\tablecomments{Table \ref{tab7} is published in its entirety 
in the electronic edition of the {\it Astrophysical Journal}.
A portion is shown here for guidance regarding its form and content.}
\end{deluxetable}

%% file: tab8.tex
%\documentclass[12pt,preprint]{aastex}
%\begin{document}

\begin{deluxetable}{lr}
\tabletypesize{\scriptsize}
%\rotate
\tablecolumns{2}
\tablenum{8}
\tablecaption{Number of sources in the IRAC Galactic Center Catalog.
\label{tab8}}
\tablewidth{0pc}
\tablehead{
\colhead{Quality} & 
\colhead{Number of Sources} }
\startdata
Whole Catalog                            & 1,065,565 \\
``2+1" Flag = `1'                        &   656,673 \\
Pos. Flag = `1' (ok)                     &   938,681 \\
Pos. Flag = `2' (Central Cluster)        &       104 \\
Pos. Flag = `3' (Quintuplet Cluster)     &        90 \\
Pos. Flag = `0' (Incomplete Coverage)    &   126,690 \\ \hline
\multicolumn{2}{c}{IRAC Channel 1}                   \\ \hline
Detected Sources                         &   735,011 \\
Sources with ``2+1" Flag = `1', SNR$>$10 &   484,810 \\
PRF fluxes                               &   711,926 \\
Aperture fluxes                          &    10,047 \\
Sub-array fluxes                         &       177 \\
Saturated fluxes                         &    12,861 \\ \hline
\multicolumn{2}{c}{IRAC Channel 2}                   \\ \hline
Detected Sources                         &   700,918 \\
Sources with ``2+1" Flag = `1', SNR$>$10 &   449,496 \\
PRF fluxes                               &   682,367 \\
Aperture fluxes                          &    11,354 \\
Sub-array fluxes                         &       149 \\
Saturated fluxes                         &     7,048 \\ \hline
\multicolumn{2}{c}{IRAC Channel 3}                   \\ \hline
Detected Sources                         &   493,190 \\
Sources with ``2+1" Flag = `1', SNR$>$10 &   343,893 \\
PRF fluxes                               &   477,152 \\
Aperture fluxes                          &    15,430 \\
Sub-array fluxes                         &        82 \\
Saturated fluxes                         &       526 \\ \hline
\multicolumn{2}{c}{IRAC Channel 4}                   \\ \hline
Detected Sources                         &   323,514 \\
Sources with ``2+1" Flag = `1', SNR$>$10 &   200,167 \\
PRF fluxes                               &   310,436 \\
Aperture fluxes                          &    12,247 \\
Sub-array fluxes                         &        57 \\
Saturated fluxes                         &       774 \\
\enddata 
\end{deluxetable}

%\end{document}